\documentclass{jfm}
\usepackage[utf8]{inputenc}
\usepackage[letterpaper, left=2.0cm, right=2.0cm, top=2.0cm]{geometry}
\usepackage{lipsum}
\usepackage{xcolor}
\usepackage{mathtools}
\usepackage{natbib}
\usepackage{amsmath}
\usepackage{soul}
\usepackage{wasysym}
\usepackage{braket}
\usepackage{subfig}
\usepackage[hidelinks,colorlinks=true,linkcolor=blue,citecolor=blue]{hyperref}

\date{}
\usepackage{tikz}
\usetikzlibrary{shapes}
\usetikzlibrary{plotmarks}
\usepackage{graphicx}
\usepackage{multirow}

\newcommand{\blackrec}{\raisebox{0pt}{\tikz{\draw[black,solid,line width = 1.0pt](2.mm,0) rectangle (3.5mm,1.5mm);\draw[-,black,solid,line width = 1.0pt](0.,0.8mm) -- (5.5mm,0.8mm)}}}
\newcommand{\blackcircle}{\raisebox{0pt}{\tikz{\draw[black,solid,line width = 1.0pt](3.mm,0) circle (1.0mm);\draw[-,black,solid,line width = 1.0pt](0.,0.0mm) -- (5.5mm,0.0mm)}}}

\newcommand{\blackut}{\raisebox{0pt}{\tikz{\node[draw,scale=0.4,regular polygon, regular polygon sides=3,rotate=180,xshift=-7mm,line width = 1.0pt](){};\draw[-,black,solid,line width = 1.0pt](0.,0.0mm) -- (5.5mm,0.0mm)}}}

\newcommand{\blackdt}{\raisebox{0pt}{\tikz{\node[draw,scale=0.4,regular polygon, regular polygon sides=3,rotate=0,xshift=7mm,line width = 1.0pt](){};\draw[-,black,solid,line width = 1.0pt](0.,0.0mm) -- (5.5mm,0.0mm)}}}

\newcommand{\blackst}{\raisebox{0pt}{\tikz{\node[star,star point ratio=2.25,minimum size=7pt,inner sep=0pt,draw=black,solid,xshift=3mm,line width = 1.0pt] {};\draw[-,black,solid,line width = 1.0pt](0.,0.0mm) -- (5.5mm,0.0mm)}}}

\begin{document}

\shorttitle{DNS of turbulent pipe flow up to $Re_\tau=5200$ } 
\shortauthor{Yao, Rezaeiravesh, Schlatter and Hussain} 
\title{Direct numerical simulation of turbulent pipe flow up to $Re_\tau\approx5200$}

\author{
 Jie Yao\aff{1},
Saleh Rezaeiravesh\aff{2},
 Philipp Schlatter\aff{2} and  
 Fazle Hussain\aff{1},
 }
\affiliation{
\aff{1}
Department of Mechanical Engineering, Texas Tech University,  Lubbock, Texas, USA, 79409

\aff{2}
 Department of Engineering Mechanics,	KTH Royal Institute of Technology, SE-100 44 Stockholm, Sweden
}

\maketitle

\begin{abstract}
Well-resolved direct numerical simulations (DNSs) have been performed of the flow in a smooth circular pipe of radius $R$ and   axial length $10\pi R$ at friction Reynolds numbers up to $Re_\tau=5200$ using the pseudospectral code OPENPIPEFLOW. Various turbulence statistics are documented and compared with other DNS and experimental data  in pipes as well as channels.
Small but distinct differences  between various datasets are identified.
The friction factor $\lambda$ overshoots by $2\%$  and undershoots by $0.6\%$ of the Prandtl friction law  at low and high $Re$ ranges, respectively. 
In addition, $\lambda$  in our results is slightly higher  than that in \citeauthor{pirozzoli2021one} (\citeauthor{pirozzoli2021one}, J. Fluid. Mech., 926, A28, 2021), 
but matches well with the experiments in \citeauthor{furuichi2015friction} (\citeauthor{furuichi2015friction}, Phys. Fluids, 27, 095108, 2015).
The log-law indicator function, which  is nearly indistinguishable between the pipe and channel flows up to $y^+=250$,   has not yet developed a plateau further away from the wall in the pipes even for the  $Re_\tau=5200$ cases.
The  wall shear stress fluctuations and the inner peak of the axial velocity intensity -- which grow monotonically with $Re_\tau$ -- are  lower in the pipe than in the channel, but the difference  decreases with increasing $Re_\tau$.
While the wall values  are slightly lower in channel than pipe flows at the same $Re_\tau$, the inner peaks of  the pressure fluctuations show negligible differences between them.
The Reynolds number scaling of all these quantities agrees with both the logarithmic  and defect power laws if the  coefficients are properly chosen.
 The one-dimensional spectrum of the axial velocity fluctuation exhibits a $k^{-1}$ dependence at an intermediate distance from the wall -- as also seen   in the channel flow. 
In summary, this high-fidelity data enable us to provide better insights into the flow physics in the pipes as well as the similarity/difference among different types of wall turbulence. 
\end{abstract}
%
%
\section{Introduction}
Turbulent flows that are constrained by a wall (referred to as ``wall turbulence'') are common in nature and engineering applications. 
Roughly half of the energy spent  in transporting fluids through pipes or vehicles through air or water is dissipated by the turbulence near the walls \citep{jimenez2012cascades}. 
Therefore, an improved understanding of the underlying physics of these flows is  essential for modeling and control  \citep{kim2011physics,canton2016large,yao2018drag}. 
The spatially evolving boundary layer, the (plane) channel, and the pipe are three canonical geometrical configurations of wall turbulence.
Different from    boundary layer  and channel flows, azimuthal periodicity is inherent to   pipe flows. 
Therefore, pipe flow is the most canonical case, being completely described by the Reynolds number ($Re$) and the axial length -- the effect of  the latter is limited if  sufficiently large \citep{el2013direct, feldmann2018computational}.


The sustained interest in high-$Re$ wall turbulence stems from numerous  open questions regarding the scaling of turbulent statistics, as reviewed
in \cite{marusic2010wall} and \cite{smits2011high}. 
For example, a  characteristic of high-$Re$  wall turbulence is the logarithmic law in the mean  velocity with an important parameter, {i.e.} the von K\'{a}rman constant $\kappa$, whose value and  universality among different flow geometries  are still highly debated \citep{nagib2008variations,she2017quantifying}.
Also, no conclusion has been reached on whether the near-wall peak of the streamwise velocity fluctuations   continuously increases with $Re$ \citep{marusic2017PRF} or eventually saturates at high $Re$ \citep{chen2021reynolds,klewicki_2022}.
Furthermore,  the existence of  an outer peak in the streamwise velocity fluctuations, as indicated by experiments,  is also highly debated \citep{Hultmark2012PRL,willert2017near}. 
Other questions, which can  be answered only with substantially higher $Re$'s, are  the scaling and   generation mechanism of the various flow structures.
Large-scale (LSMs) and very large-scale motions (VLSMs), with lengths of $5R$ up to $20R$ have been experimentally found in the outer region of  pipe flows \citep{kim1999very,monty2007large}. 
Here, $R$ is the radius of the pipe. 
Due to the increasing strength of these structures with $Re$, they even leave their footprint quite close to the wall \citep{monty2007large}, in the form of amplitude modulation as reviewed by {e.g.}\ \cite{dogan2018quantification}. 

Fundamental studies of  wall turbulence  require accurate representations or measurements of the flows, which were typically carried out via  experiments \citep{zagarola1998,mckeon2005new,smits2011spatial,furuichi2015friction,furuichi2018further,talamelli2009ciclope,fiorini2017turbulent}. 
However, decades of experimental research  have shown that  obtaining unambiguous high-$Re$   data, particularly   near the wall, remains a challenge.
That is because as  $Re$ increases,  the smallest scale decrease -- leading to large uncertainties in  determining the probe locations and  turbulence intensities.
Advances in computer technology (both in  hardware and software) have enshrined direct numerical simulation (DNS) as an essential tool for turbulence research.
Although only moderate $Re$ can be achieved at the current stage, DNS  provides extensive, detailed data  compared to experiments -- even close to the walls where experimental data is very difficult to be obtained.
One of the earliest DNS for wall turbulence was performed by \cite{kim1987turbulence} for the channel flow at friction Reynolds number  $Re_\tau (\equiv u_\tau h/\nu)\approx180$ (here, $u_\tau$ is the friction velocity, $h$ is the half channel height, and $\nu$ is the fluid kinematic viscosity).
They found good agreements between DNS and  experimental data  by \cite{hussain1975measurements}, except in the near-wall region.
The discrepancy was speculated to be  caused by
the inherent near-wall turbulence  measurement  errors.
\cite{eggels1994fully} subsequently conducted the first DNS of pipe flow  at $Re_\tau (\equiv u_\tau R/\nu) \approx180$ to investigate the differences between channel and pipe flows.

Numerous DNS investigations have been carried out in the aftermath of these pioneering studies, with $Re$ progressively increasing  as a result of increased computational power \citep{moser1999direct,bernardini2014velocity,wu2008direct}. 
However, among them, only those with $Re_\tau \ge 10^3$  are of particular engineering interest as this is the range of $Re$ relevant to industrial applications. In addition, it is in this range that the high-$Re$ characteristics of wall turbulence  start to manifest. 
One of the highest $Re_\tau$ large domain  DNS  was performed by  \cite{lee2015direct} for  channel flows at $Re_\tau = 5186$ with the  domain size $L_x\times L_z=8\pi h\times3\pi h$. 
Compared to numerous  DNS  for channel flows \citep{lozano2014effect,bernardini2014velocity,lee2015direct}, fewer high-$Re$ studies have been performed for pipe flow; and most of them are limited to  $Re_\tau\approx 1000$. For example,  \cite{lee2013comparison} performed DNS at $Re_\tau\approx1000$ with a   length  of $30R$ and  established the existence of VLSM of scale up to $\mathcal{O}(20R)$. 
\cite{el2013direct}  used a spectral-element method  to perform DNS for $Re_\tau$ up to 1000 with the length $L_z=25R$.
\cite{chin2014reynolds} found that the mean velocity profile does not exhibit a strictly logarithmic  layer with $Re_\tau$ up to 2000, necessitating a finite-$Re$ correction like those introduced by  \cite{afzal1976millikan} and \cite{jimenez2007we}.
To quantify the effects of computational length and $Re$,  \cite{feldmann2018computational} conducted DNS for $90\le Re_\tau\le 1500$ with  $L_z$ up to $42R$. 
They confirmed that $L_z=42R$ is sufficiently large to capture the LSM and VLSM-relevant  scales.
\cite{ahn2015direct} performed DNS of pipe flow at $Re_\tau\approx3000$  for a  length of $30R$.
They claimed that the streamwise mean velocity profiles followed a power law in the overlap region and observed a clear scale separation between inner- and outer-scale turbulence.
So far, the largest DNS of pipe flow  is done  at $Re_\tau\approx6000$  with a relatively short length ($L_z=15R$) by \cite{pirozzoli2021one} based on a lower-order numerical method.

In general, one would anticipate that various simulations and experiments  to agree  with each other to a high degree.
However, a comparison among several   datasets in spatially developing turbulent boundary layers \citep{schlatter2010assessment}, channels \citep{lee2015direct}, and pipes \citep{pirozzoli2021one} flows surprisingly shows considerable variations among  the various DNSs, even for basic measures such as the shape factor, friction coefficient,  the von K\'{a}rman constant, etc. 
Accurate turbulence statistics are very much needed, both for  understanding turbulence physics and for developing, adapting and validating  turbulence models.
Here, we present a new high-fidelity DNS  dataset of turbulent pipe flow generated with a pseudo-spectral method for  $Re_\tau$ up to $5200$ and with the axial length  $L_z/R=10\pi$, which is long enough to capture the LSMs and VLSMs  reported in experimental studies \citep{guala2006large}.
The accuracy of this  dataset is quantified by using the newly developed uncertainty quantification method.
In addition, the dataset is extensively compared with other DNS and experimental data  for turbulent pipe and channel flows.

\section{Simulation details}

\begin{table}
\centering
        \begin{tabular}{cccccccccc}
			$Re_\tau$ & $Re_b$ & $N_z\times N_r\times N_\theta$&$\Delta z^+$&$\Delta r^+_w$&$\Delta r^+_{\max}$&$\Delta (R\theta)^+$&$Tu_\tau/R$&Symbols\\
            $181$& $5300$  & $1024 \times 192\times256$& $5.5$&$0.02$&$2.0$&$4.4$&105.3&\protect\blackrec\\
             $549$& $19,000$ & $2048 \times 256\times768$& $8.4$&$0.09$&$3.2$&$4.5$&24.7&\protect\blackcircle\\
             $998$& $37700$  & $3072 \times 384\times1280$& $10.2$&$0.1$& $3.9$&$4.9$&17.1&\protect\blackut\\
			$2001$& $83000$ & $6144 \times 768\times2560$& $10.2$&$0.1$&$3.9$&$4.9$&9.7&\protect\blackdt\\
			$5197$& $240000$ & $12288 \times 1024\times5120$& $12.8$&$0.2$&$8.6$&$6.3$&4.6&\protect\blackst\\
        \end{tabular}
         \caption{Summary of simulation parameters. The axial length of the pipe  ($L_z$) is $10\pi R$, with $R$ being the pipe radius.
        Here, $Re_\tau (\equiv u_\tau R/\nu)$ and $Re_b (\equiv 2U_b R/\nu)$ are the frictional and bulk Reynolds numbers, respectively. $N_z$ and $N_\theta$ are the number of   dealiased Fourier modes in axial and azimuthal directions, and $N_r$ are the number of grid points in the radial direction.  $\Delta z$ and $\Delta (R\theta)$ are the grid spacing in the axial and azimuthal directions, defined in terms of the Fourier modes. In the  radial direction, $\Delta r^+_w$ represents the grid spacing at the wall, and $\Delta r^+_{max}$ denotes the maximum grid spacing. $T u_\tau/R$ is the total eddy-turnover times without  the initial transient phase.}{\label{tbl:Num1}}
\end{table}

DNSs of incompressible turbulent pipe flows are performed using the pseudo-spectral code ``OPENPIPEFLOW'' developed  by  \cite{willis2017openpipeflow}.
The radial, axial, and azimuthal directions are represented by $r$, $z$, and $\theta$, and the corresponding velocity components are $u_r$, $u_z$, and $u_\theta$.
Fourier discretization  is employed in the periodic axial ($z$) and azimuthal ($\theta$) directions,   while a central finite difference scheme with a nine-point stencil is adopted in the radial ($r$) direction. 
The number of grid points in $r$--direction is $N_r$, and the number of  Fourier modes in  $z$-- and $\theta$-- directions are $N_z$ and $N_\theta$, respectively.
In the physical space, the
number of grid points in the $z$-- and $\theta$-- directions
increases by a factor of 3/2 due to dealiasing.
Grid points are distributed in $r$--direction according to a hyperbolic tangent function so that high wall-normal velocity gradients in the viscous sublayer can be resolved. In addition,
the first few points near $r = 0$ are also clustered to preserve the high order of the finite difference scheme across the pipe axis.
The governing equations are integrated with a second-order semi-implicit time-stepping scheme. 
The flow is driven by a  pressure gradient, which varies in time to ensure that the mass flux  through  the  pipe  remains  constant.
For   more   details   about   the   code   and the   numerical   methods,  see \cite{willis2017openpipeflow}.

Five different Reynolds
numbers $Re_\tau\approx 180$, $550$, $1000$, $2000$, and $5200$ are considered.
The detailed simulation parameters, such as domain sizes and grid sizes,  etc.  are listed in Table \ref{tbl:Num1}.
The simulations are performed with resolutions that are comparable to those used in the prior  simulations, e.g. \cite{lozano2014effect} and \cite{lee2015direct}.
In particular, for $Re_\tau\le2000$, the axial and azimuthal resolutions employed here satisfy the criterion suggested by \cite{yang2021grid} for capturing  $99\%$ of the  wall shear stress events.
For the highest  $Re_\tau$ case (i.e. $\approx 5200$), $N_z=12288$ and $N_\theta=5120$ Fourier modes are used in the $z$-- and $\theta$--directions -- corresponding to an effective resolution of 
$\Delta z^+=L^+_x/N_z=12.8$ and $\Delta (R\theta)^+=(2\pi R^+)/N_\theta=5.1$.
Hereinafter, the superscript $+$ indicates non-dimensionalization in wall units, i.e. with kinematic viscosity $\nu$ and friction velocity $u_\tau$. 
For comparison, several DNS and experimental data  from  the literature are included.
The details are listed in table  \ref{tbl:Num2}.
To further validate the accuracy of our simulation, an additional simulation at $Re_\tau=2000$ is performed  using NEK5000 {(hereinafter, this case is denoted as NEK5000 2K)}.
The numerical setup and mesh generation are the same as those in \cite{el2013direct}.
The length of the pipe is chosen as $L_z=35R$, and the  total number of  spectral elements is $7598080$. 
With the polynomial order set to $12$, the total number of grid points is approximately $13.1\times 10^9$. 
The grid spacing  is comparable to that used in  \cite{el2013direct} in all directions.

\begin{table}
\centering
        \begin{tabular}{cccccccccc}
			Reference& Type & $Re_\tau$ range &Method\\
		\cite{wu2008direct}&DNS& 180,1140&FD\\
		\cite{el2013direct}& DNS& $180-1000$&SE\\
		\cite{chin2014reynolds}&DNS&$180-2000$&SE\\
		\cite{ahn2013direct,ahn2015direct}& DNS& $3000$&FD\\
		\cite{pirozzoli2021one}& DNS& $180-6000$&FD\\
		\cite{furuichi2015friction,furuichi2018further}&EXP&  $1000-53000$&LDV&\\
        \cite{lee2015direct} & DNS (channel)  & $180-5200$&SB \\
        \cite{hoyas2022wall}&DNS (channel)  & $10000$&SC\\
        \end{tabular}
         \caption{List of references of data used.
     FD denotes finite difference, SE denotes spectral element, SB denotes spectral/B-spline, SC denotes spectral/compact finite difference, and LDV represents laser Doppler velocimetry.}{\label{tbl:Num2}}
\end{table}

The uncertainty in the flow quantities due to the finite time-averaging is estimated using the methods described in \cite{uqRecipes} and \cite{xavier:22}. For the central moments of velocity and pressure, the central limit theorem is applied to the time samples averaged over the $z$-- and $\theta$--directions. 
The associated time-averaging uncertainty is estimated using an autoregressive-based model for the autocorrelation function, see e.g. ~\cite{oliver:14} and \cite{xavier:22}. 
For estimating the uncertainty in the combination of central moments, the method proposed   by \cite{uqRecipes} is employed. 
See Appendix \ref{sec:appa}  for further discussion on the method and estimated uncertainties in the first- and second-order velocity moments. 
For $Re_\tau=5200$ case, the estimated standard
deviation of the mean axial velocity ($U^+$) is less than 0.1\%, and the estimated standard deviation of the velocity variance (i.e. $\braket{u'^2_r}^+$, $\braket{u'^2_\theta}^+$, $\braket{u'^2_z}^+$) and covariance ($\braket{u'_ru'_z}^+$) is less than 1\% in the near-wall region ($y^+ < 100 $) and about 5\% in the center region.
Hereinafter, the velocity fluctuations are denoted using the prime symbol (e.g. $u'_r$), and the ensemble (both in time and space) averaged quantities of the mean velocity and velocity fluctuations are expressed using a capital letter or bracket (e.g. $U$ or $\braket{u'_r u'_\theta}$).

\begin{figure}
\centering
        \includegraphics[width=0.6\textwidth]{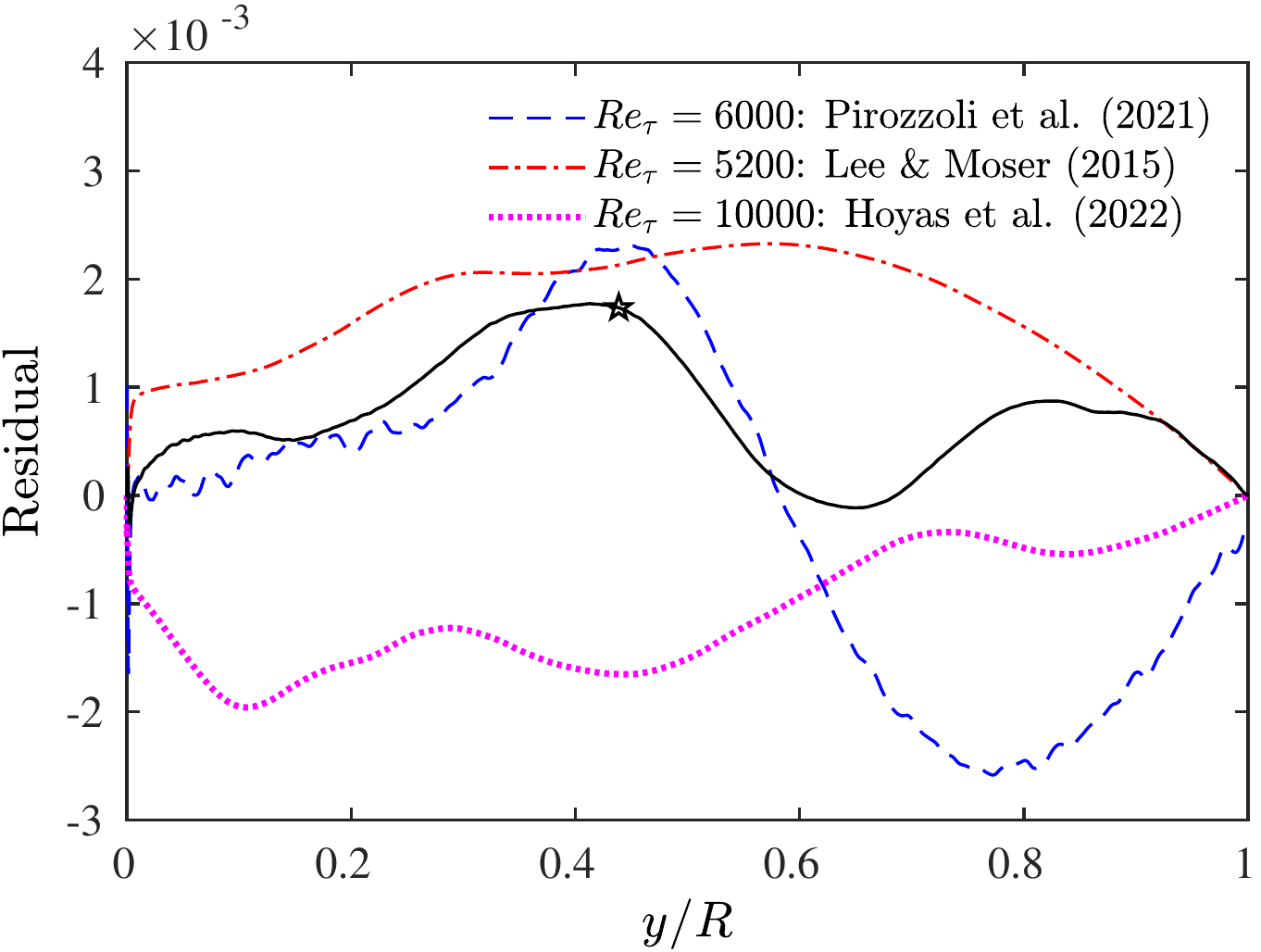}
  
    \caption{Comparison of the residual in the mean momentum equation  \eqref{eqn:moment} among different high $Re$ simulations. {The black solid line with star denotes our  $Re_\tau=5200$ result.}}
    \label{fig:resd}
\end{figure}

In addition, the mean momentum equation is employed to ensure that the simulation is statistically stationary.   
Due to  momentum balance, the total stress, which is the sum of Reynolds shear stress $\braket{u'_ru'_\theta}^+$ and mean viscous stress $\partial U^+/\partial y^+$, is linear in a statistically stationary turbulent   pipe flow:
\begin{equation} \label{eqn:moment}
    \frac{\partial U^+}{\partial y^+}-\braket{u'_ru'_\theta}^+=1-y/R,
\end{equation}
here $y=R-r$.
Figure \ref{fig:resd} shows the residual in \eqref{eqn:moment}  for the  $Re_\tau\approx5200$ case.
The discrepancy between the analytic linear profile (i.e. $1-y/R$) and total stress
profile (i.e. $\partial U^+/\partial y-\braket{u'_ru'_\theta}^+$) from the simulation is less than 0.002 in  wall units and is comparable to  other high--$Re$ DNS  in the literature. 
Note that this discrepancy is much smaller than the standard deviation of the estimated total stress (see appendix \ref{sec:appa}).

\section{Results}

\begin{table}
\centering
        \begin{tabular}{ccccccccccc}
			$Re_\tau$ & $10^2\lambda$ &  $\braket{u'^2_z}^+_p$&$-\braket{u'_r u'_z}^+_p$&$\braket{\tau'^2_{z,w} }^+$&$\braket{\tau'^2_{\theta,w} }^+$&$p'^+_{w,rms}$&$p'^+_{p,rms}$\\
            $181$& $3.730(0.20\%)$  & $7.229(0.41\%)$&$0.719 (0.34\%)$&$0.123(0.54)$&$0.031(1.18\%)$&$1.721(1.10\%)$&$2.039(0.90\%)$\\
             $549$& $2.672(0.10\%)$ & $7.587(0.14\%)$&$0.861(0.13\%)$&$0.158(0.14\%)$&$0.063(0.34\%)$&$2.283(0.62\%)$&$2.720(0.69\%)$\\
             $998$& $2.244(0.14\%)$   & $7.978(0.23\%)$&$0.902(0.13\%)$&$0.170(0.14\%)$&$0.071(0.15\%)$&2.569(0.50\%)&3.020(0.38\%)\\
			$2001$& $1.859(0.22\%)$  & $8.522(0.14\%)$&$0.932(0.14\%)$&$0.184(0.18\%)$&$0.079(0.18\%)$&2.793(0.67\%)&3.273(0.54\%)\\
			$5197$& $1.498(0.24\%)$ & $9.117(0.21\%)$&$0.957(0.12\%)$&$0.196(0.16\%)$&$0.082(0.28\%)$&3.174(0.80\%)&3.610(0.70\%)\\
        \end{tabular}
         \caption{Summary of values and  standard deviation of some key parameters: the fiction factor
         $\lambda=8\tau_{z,w}/(\rho U^2_b)$; the peak of axial velocity variance $\braket{u'^2_z}^+_p$; the peak of the Reynolds shear stress $-\braket{u'_ru'_z}^+_p$; the axial $\braket{\tau'^2_{z,w} }^+$ and azimuthal $\braket{\tau'^2_{\theta,w} }^+$ wall shear stress fluctuations; and the wall ($p'^+_{w,rms}$) and peak ($p'^+_{p,rms}$) values of root-mean-square (r.m.s.)  pressure fluctuations.}{\label{tbl:Num3}}
\end{table}

\subsection{Flow visualization}

\begin{figure}
    \centering
        \includegraphics[width=0.9\textwidth]{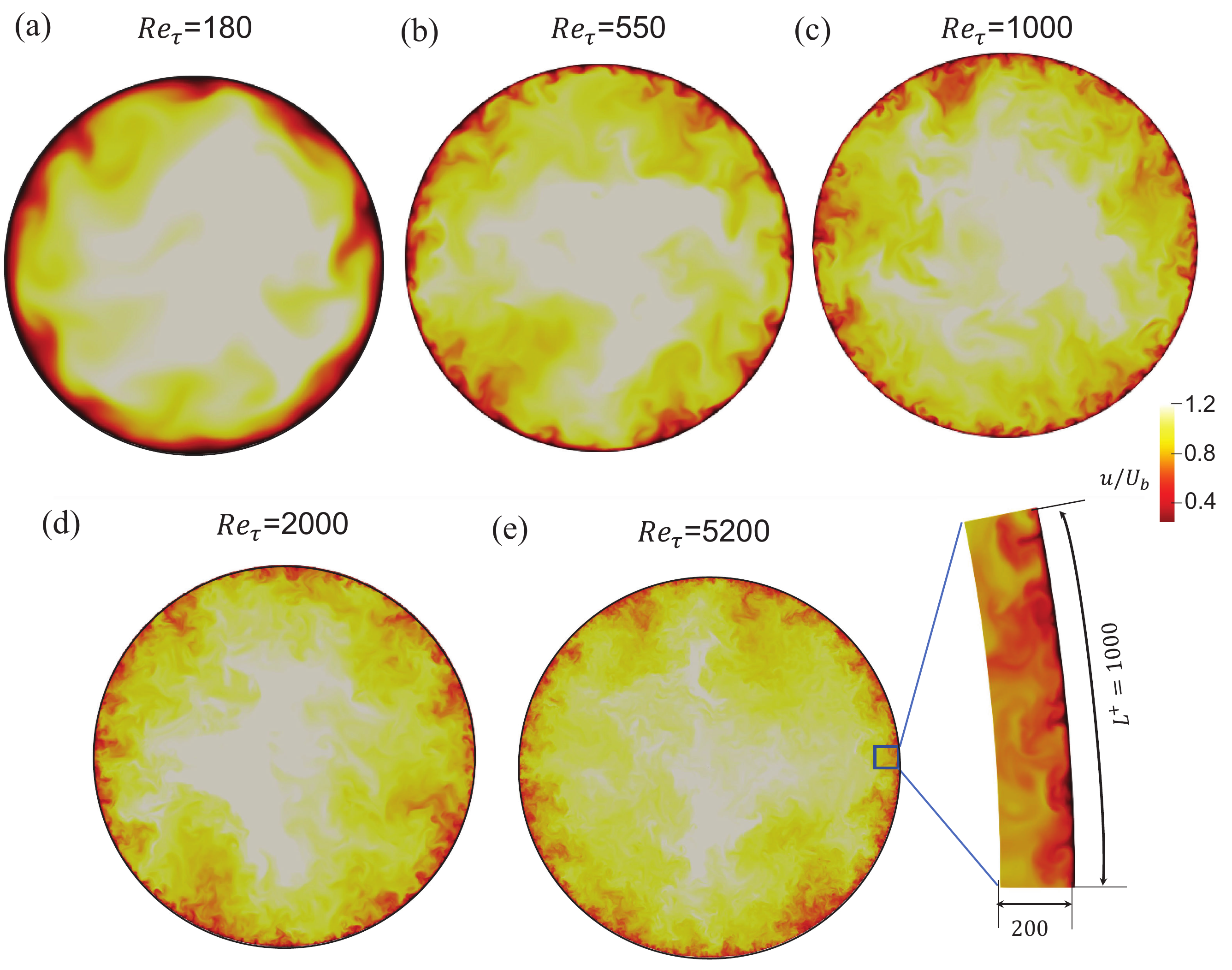}
    \caption{Visualization of the instantaneous axial velocity $u_z/U_b$ for (a) $Re_\tau=180$, (b) $550$, (c) 1000, (d) 2000, and (e) 5200.}
    \label{fig:visu}
\end{figure}

The $Re$ effect on the flow structure is qualitatively
illustrated in figure \ref{fig:visu}, showing cross-sectional views of the instantaneous  axial velocity $u_z$.
Although large scales dominate in the central region of the pipe for all $Re_\tau$ cases,  there is a  general increase in the range of scales with
increasing $Re_\tau$. 
The average spacing between near-wall low-speed streaks is around $(R\theta)^+=100$. 
For the lowest $Re_\tau$ studied here, 
  about ten evenly distributed  low-speed structures are seen in figure \ref{fig:visu}(a), identified  by the plume-shaped black regions ejecting from the  wall.
For our highest $Re_\tau$, the  streak spacing is reduced to about $0.02R$, and  these fine-scale  streaks can hardly be identified from the full cross-section  in figure \ref{fig:visu}(e). 
A zoomed-in view of the near-wall region with domain size $(1000,200)$ in wall units in $(r,\theta)$ directions is provided to better visualize these  structures, which share quite similar patterns as those in low $Re_\tau$ cases.

\subsection{Friction factor}

The mean friction (or wall-shear stress), which is proportional to the pressure drop or the amount of energy required  to sustain the flow, is an important parameter and  has been extensively studied \citep{blasius1913aehnlichkeitsgesetz,mckeon2005new,furuichi2015friction,pirozzoli2021one}.
A semi-empirical relation between the friction factor $\lambda=8\tau_{z,w}/(\rho U^2_b)$ and $Re$ is given as (known as  the Prandtl friction law):
\begin{equation}\label{eqn:prand}
    1/\lambda^{1/2}=A\log_{10}(Re_b \lambda^{1/2})-B,
\end{equation}
where the constant $A$ is related  to the von   K\'{a}rman constant  as $A=1/(2\kappa \sqrt{2} \log_{10}(e))$.
Curve-fitting the experimental data  over $3.1\times 10^3<Re_b<3.2\times 10^6$ by \cite{nikuradse1933stromungsgesetze} yields  $A=2.0$ and $B=0.8$, which corresponds to $\kappa=0.407$.
However, notable deviations were observed when comparing the Prandtl friction law with other experimental data. For example, \cite{mckeon2005new} showed that for the Princeton Superpipe data (in the range of $3.1\times 10^4 \leq Re_b \leq 3.5 \times 10^7$), the constants of the Prandtl law  work only over a limited range of  $Re_b$. New constants (i.e. $A=1.920$ and $B=0.475$) and additional $Re$-dependent corrections need to be introduced  to better fit  the data in the entire $Re_b$ range.
For the ``Hi-Reff'' data, \cite{furuichi2015friction} found that $\lambda$  deviates  from the Prandtl  law with  approximately $2.5\%$ in the lower $Re_b$ and $-3\%$ in the high $Re_b$ region.
In addition, $\lambda$, although agreeing with the Superpipe data in the low $Re_b$ range, deviates for $Re_b>2 \times 10^5$.

\begin{figure}
    \centering

    \subfloat[]{
        \includegraphics[width=0.48\textwidth]{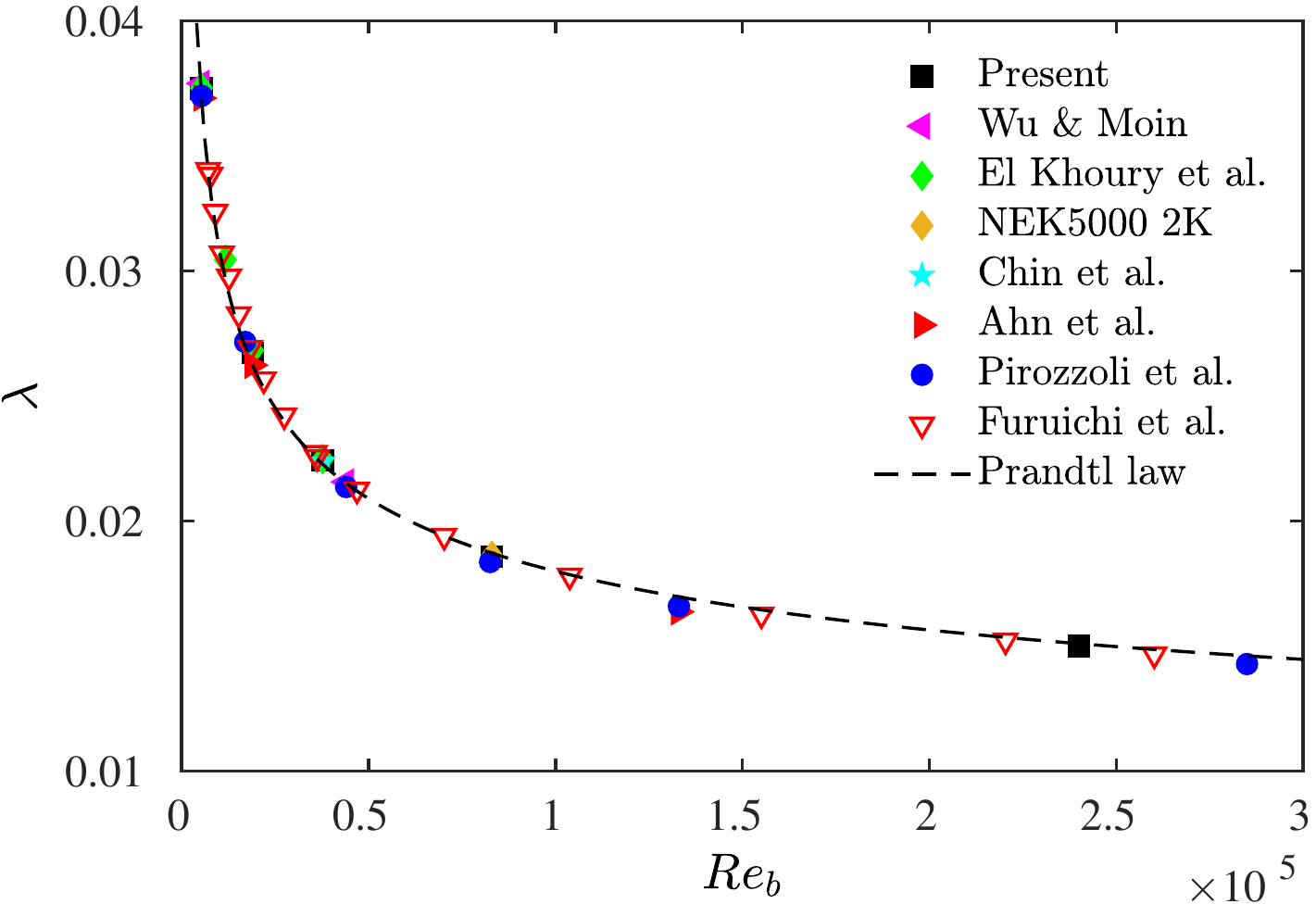}}
    \subfloat[]{
        \includegraphics[width=0.48\textwidth]{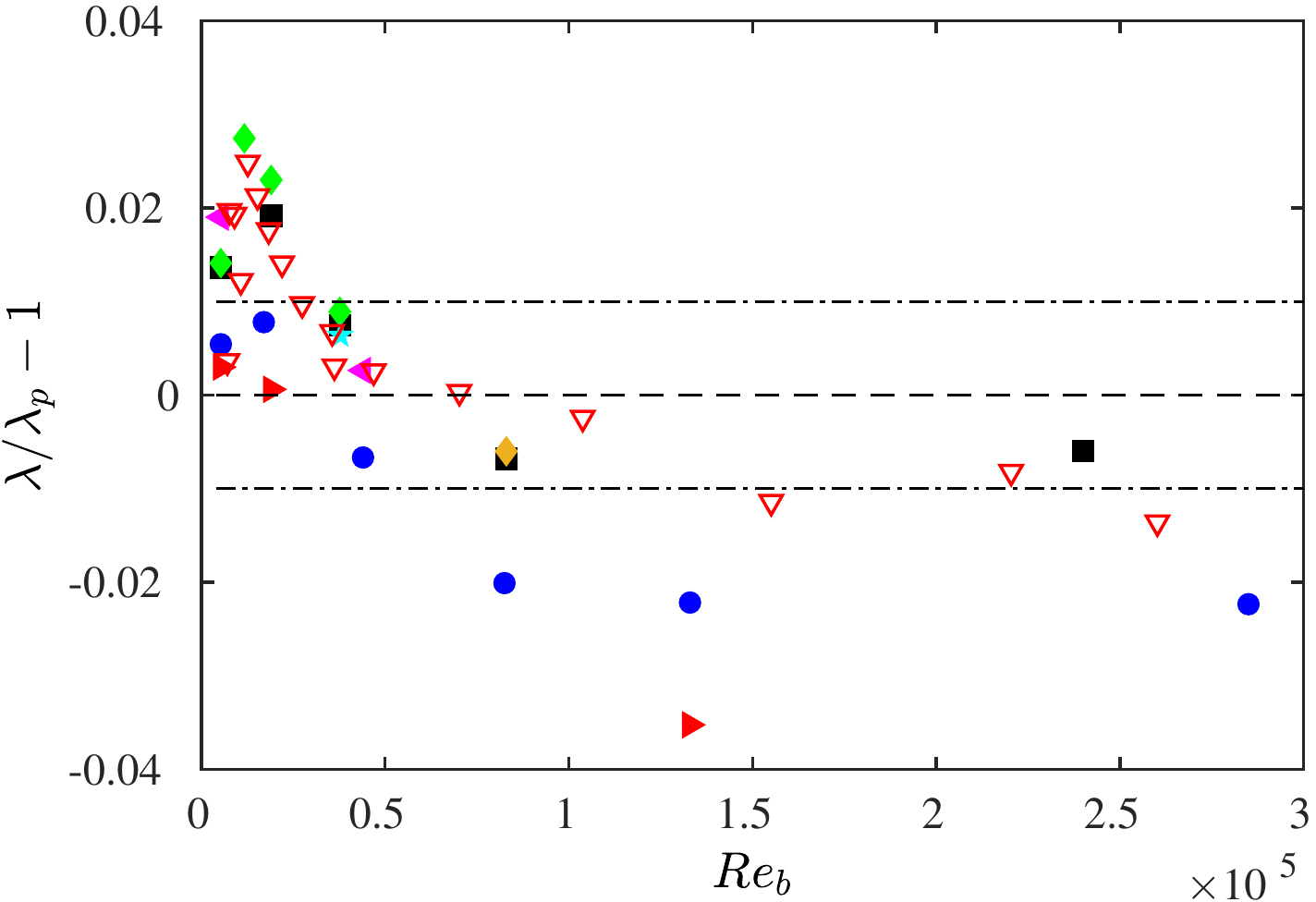}}\\
    \caption{ Friction factor $\lambda$ as a function of  $Re_b$ and (b) the relative deviations from the Prandtl friction law.}
        \label{fig:lam}
\end{figure}

Figure \ref{fig:lam}(a) shows the friction factor $\lambda$  as a function of $Re_b$, along with other DNS and experimental data as well as the  theoretical prediction $\lambda_p$ based on  \eqref{eqn:prand}.  
All DNS and experimental data seem to follow~$\lambda_p$. 
However, the scatter is better highlighted  by examining the relative error with respect to  the Prandtl law (i.e. $\lambda/\lambda_p-1$).
As depicted in   figure \ref{fig:lam}(b),
all DNS data overshoot $\lambda_p$ at the low $Re_b$  (i.e. $\le4\times 10^4$).
{Our data, which agrees with \cite{wu2008direct,el2013direct,chin2014reynolds} and the new simulation at $Re=2000$ using NEK5000}, exceed $\lambda_p$ by about 2\%, while the results of \cite{ahn2013direct} and \cite{pirozzoli2021one} are closer to $\lambda_p$ (within 1\% for $Re_b<5\times 10^4$).
\cite{pirozzoli2021one} attributed this discrepancy to the different grid resolutions employed in the $\theta$-direction, which is $(R\theta)^+=4-5$ in theirs and Ref. \cite{ahn2013direct}, but $7-8$ for  \cite{wu2008direct} and \cite{chin2014reynolds}. 
However, the data from \cite{el2013direct} and our DNS have comparable azimuthal resolutions as  \cite{ahn2013direct} and \cite{pirozzoli2021one} but produce  results similar to \cite{wu2008direct} and \cite{chin2014reynolds} -- suggesting that the azimuthal resolution is not the main reason for such discrepancy. 
Interestingly,  the data of \cite{ahn2013direct} and \cite{pirozzoli2021one} are  consistently lower than our and other results in the whole $Re_b$ range. 
In particular, at the highest $Re_b$, \cite{pirozzoli2021one}'s data undershoots by~2\% from $\lambda_p$, but our data is lower  by only  about 0.6\%. 
Table \ref{tbl:Num3} asserts that such  differences in $\lambda$ are beyond the uncertainty limit. 
We speculate that the discrepancy  is due to the numerical methods used --   second-order finite difference  for  \cite{ahn2013direct} and \cite{pirozzoli2021one}, and higher-order methods for others.

Our DNS and the experimental data by  \cite{furuichi2015friction} agree well for the  $Re_b$ range studied. 
Fitting our DNS data with \eqref{eqn:prand} yields $A=2.039\pm0.083$, $B=0.948\pm0.364$ with uncertainty estimates based on 95\% confidence bounds, giving $\kappa=0.399\pm0.015$.
The value reported by \cite{pirozzoli2021one} (i.e. $A=2.102$, $B=1.148$) are slightly larger than ours but still within the uncertainty range.
However,    large uncertainty is present in the fitted values due to the limited data points in $Re$. 
In addition,  as the  $Re_b$  is still relatively low, the reported value of  $\kappa=0.399\pm0.015$ should not be used outside of the given Reynolds number range. As will be shown in  section \ref{sec:menan}, even for the highest $Re_b$ case, a distinct logarithmic region does not manifest itself in  the mean velocity profile $U(y)$. 
Higher $Re$ data is required to better estimate the  constants in \eqref{eqn:prand} and the associated $\kappa$ values.

\subsection{Wall shear stress fluctuations}

The $Re$-dependence of axial wall-shear stress fluctuation  $\langle \tau'^2_{z,w} \rangle^+$ is one of the  highly debated issues in wall turbulence. 
{Note that $\langle \tau'^2_{z,w} \rangle^+$ is also equivalent  to the wall dissipation of the axial Reynolds stress components $\epsilon^+_{z,w}$,  the azimuthal vorticity variance at the wall $\braket{\omega'^2_\theta}^+$ or the limiting value  of  $\langle u'^2_z  \rangle^+/U^2$ at the wall \citep{orlu2011fluctuating}. 
Previous DNS studies on channel, pipe, and turbulent boundary layer observed an increase in $\langle \tau'^2_{z,w} \rangle^+$  with $Re_\tau$, which reflects the increased contribution of large-scale motions on wall shear stress at high $Re$'s \citep{marusic2010predictive}.
However, the exact dependence of $\langle \tau'^2_{z,w} \rangle^+$ on $Re_\tau$ is not well established. 
For example,  
\cite{orlu2011fluctuating} suggested that the r.m.s. $\langle \tau'^2_{z,w} \rangle^+$ follows
\begin{eqnarray}\label{eqn:wallog1}
\langle \tau'^2_{z,w} \rangle^{+1/2}=C+D\ln(Re_\tau),
\end{eqnarray}
where the two constants $C$ and $D$ are  chosen as $0.298$ and $0.018$ based on the DNS of turbulent boundary layer data.}

Some works \citep{yang2017multifractal,smits2021reynolds} also suggested that 
\begin{eqnarray}\label{eqn:wallog2}
\langle \tau'^2_{z,w} \rangle^+=E+F\ln(Re_\tau),
\end{eqnarray}
By fitting turbulent channel flow data of \cite{lee2015direct} and pipe flow data of \cite{pirozzoli2021one} for $Re_\tau\ge1000$, \cite{smits2021reynolds} obtained $E=0.08$ and $F=0.0139$.

Recently, \cite{chen2021reynolds} proposed  a defect power law,  given as 
\begin{eqnarray} \label{eqn:dissp}
\langle \tau'^2_{z,w} \rangle^+=\epsilon^+_{z,w}=G-H Re^{-1/4}_\tau,
\end{eqnarray}
where $G$ is the asymptotic value at infinite $Re$, and $H$ is the coefficient. 
The assumption for \eqref{eqn:dissp} is that 
the energy dissipation balances the turbulent kinetic energy production $P_k=-\braket{u_ru_z}^+(\partial U^+/\partial y^+)$ near the location of peak production. 
The fact that   $P_k$ is bounded  by 1/4 \citep{sreenivasan1989turbulent,pope2000turbulent}  implies that the  wall dissipation  $\epsilon^+_{z,w}$  may also stay bounded, which is further assumed by \cite{chen2021reynolds} to be the same bound as  $P_k$, i.e. $G=1/4$ \citep{chen2021reynolds}. 
This argument was later criticized by \cite{smits2021reynolds} for the following two reasons.
First, based on the DNS data,  the location of peak production is actually the place where the largest imbalance of  production and dissipation  occurs. 
Second,  as  the balance between different terms in the Reynolds stress transport equation rapidly changes  near the wall, it is unclear how the balance between $P_k$ and $\epsilon^+_{z,w}$ can be extended up to the  wall, where $P_k=0$, and $\epsilon^+_{z,w}$ equals  the viscous diffusion.

\begin{figure}
    \centering
        \subfloat[]{
        \includegraphics[width=0.48\textwidth]{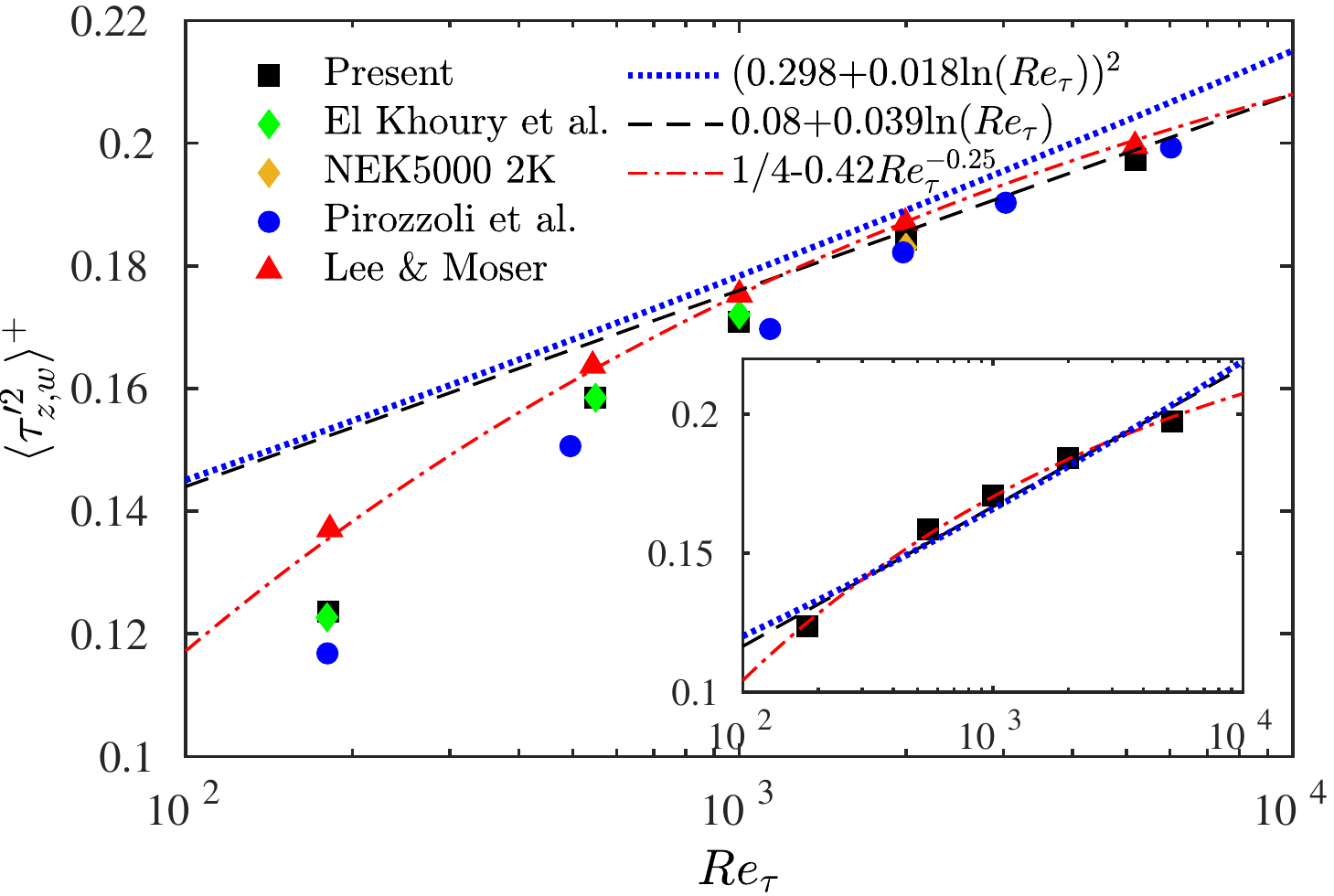}}
    \subfloat[]{
        \includegraphics[width=0.48\textwidth]{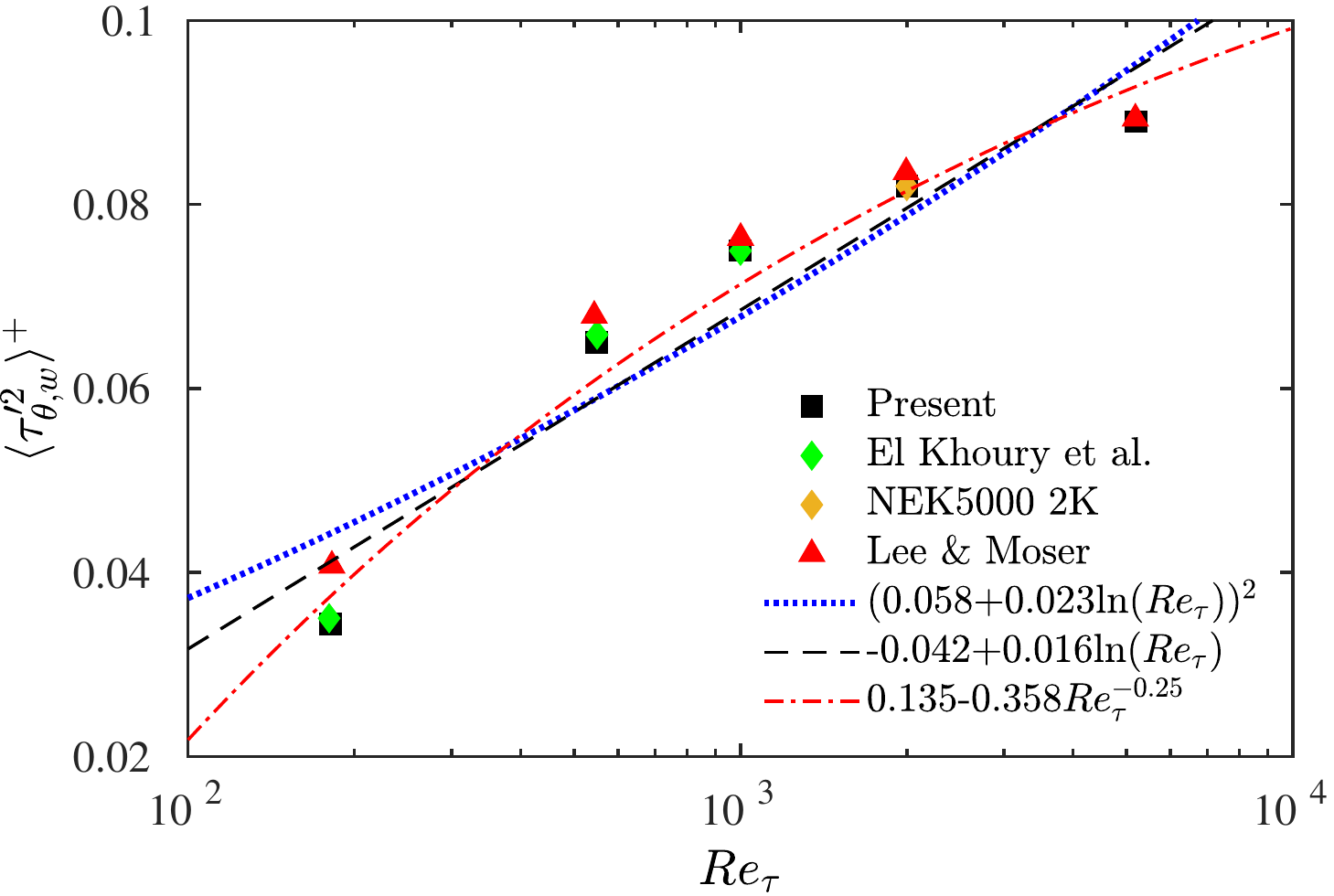}}
    \caption{(a) Axial ($\langle  \tau'^2_{z,w}\rangle^+$)  and (b) azimuthal ($\langle  \tau'^2_{\theta,w}\rangle^+$) wall shear stress fluctuations   as a function of  $Re_\tau$.
    The dotted, dashed and dashed-dotted lines in the inset of (a) denote $\langle  \tau'^2_{z,w}\rangle^+=(0.225+0.0264 \ln(Re_\tau))^2$, $\langle  \tau'^2_{z,w}\rangle^+=0.016+0.0218 \ln(Re_\tau)$ and $\langle  \tau'^2_{z,w}\rangle^+=0.255-0.477 Re^{-1/4}_\tau$, respectively.  }
    \label{fig:walls}
\end{figure}

 The axial wall-shear stress fluctuation $\langle \tau'^2_{z,w} \rangle^+$ as a function of $Re_\tau$ is depicted in figure \ref{fig:walls}(a). 
Akin to the observation in \cite{el2013direct},  the values for the pipe  are slightly lower than those for the channel, but the difference decreases with increasing $Re_\tau$.
Note that  the data of \cite{pirozzoli2021one}  is slightly lower than others, which is consistent with the lower $\lambda$ in figure \ref{fig:lam}.
The logarithmic law (\ref{eqn:wallog1}) proposed by   \cite{orlu2011fluctuating}  is higher than the DNS data. 
This is somehow expected  as the fitting coefficients were obtained based on the DNS of turbulent boundary layer data, which is higher than pipe and channels. 
In the high $Re_\tau$ range, both the logarithmic  (\ref{eqn:wallog2})   and defect power law scaling (\ref{eqn:dissp}) agree well with the data.  However, both scalings exhibit notable disagreements in the low $Re_\tau$ range.
The discrepancy seems not particularly surprising, given that the parameters in these equations are obtained from different datasets and different $Re_\tau$ ranges. 
As a  reference, the inset in figure \ref{fig:walls}(a) shows the fitting results of \eqref{eqn:wallog1} -- \eqref{eqn:dissp} using our DNS data only. 
The fitted values are $C=0.225$, $D=0.0264$ for  \eqref{eqn:wallog1}, $E=0.016$, $F=0.0218$ for  \eqref{eqn:wallog2} and $G=0.255$, $H=0.477$ for  \eqref{eqn:dissp}. 
For the $Re_\tau$ range studied, the data seem to  match better  the defect power law but with a slightly higher asymptotic value than suggested by \cite{chen2021reynolds}.
Additional data at even higher $Re_\tau$ is needed to confirm this finding.

Figure \ref{fig:walls}(b) further shows the azimuthal wall shear stress fluctuation $\langle \tau'^2_{\theta,w} \rangle^+$ as a function of $Re_\tau$. Similar to that found for $\langle \tau'^2_{z,w} \rangle^+$, our data agree well with \cite{el2013direct} and NEK5000 2K cases, and all of them become closer to  \cite{lee2015direct} with increasing $Re_\tau$.
Fitting data with the logarithmic and defect power law yields $\langle \tau'^2_{\theta,w} \rangle^+=(0.058+0.023\ln(Re_\tau))^2$,
$\langle \tau'^2_{\theta,w} \rangle^+=-0.040+0.016\ln(Re_\tau)$ and $\langle \tau'^2_{\theta,w} \rangle^+=0.135-0.353Re^{-1/4}_\tau$.
Again, the defect power law seems to  match better the DNS data, but the agreement is not as good as  for $\langle \tau'^2_{z,w} \rangle^+$.

\subsection{Mean velocity profile}\label{sec:menan}

In an overlap region between the inner and outer flows, there is  a logarithmic variation of the mean axial velocity $U^+$ profile, which is given as
\begin{eqnarray}\label{eqn:log}
U^+=\frac{1}{\kappa} \ln y^++B.
\end{eqnarray}
In a true log layer, the indicator function $\beta=y^+( \partial U^+/\partial y^+)$ is constant and equals  $1/\kappa$.

\begin{figure}
    \centering

    \subfloat[]{
        \includegraphics[width=0.48\textwidth]{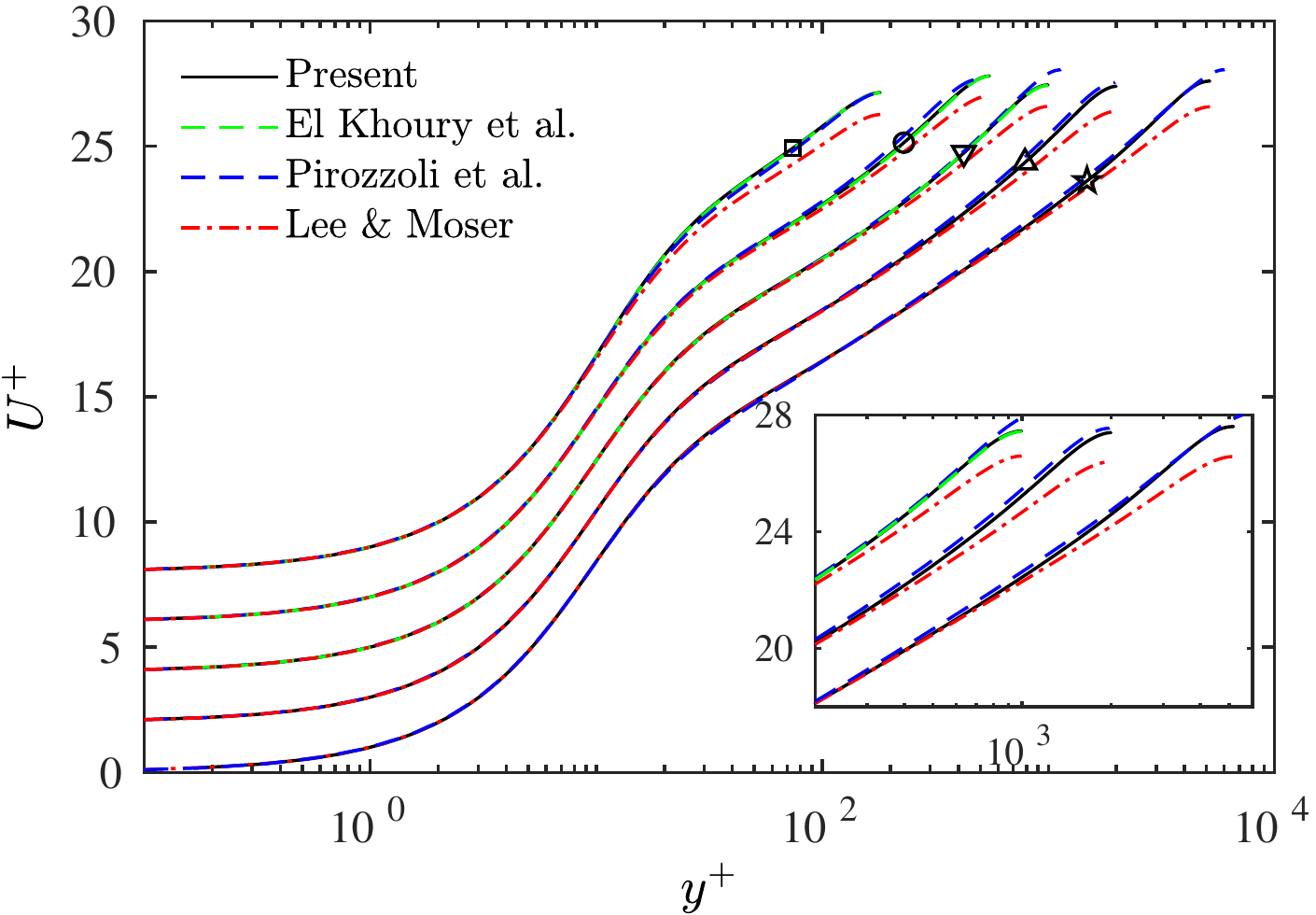}}
    \subfloat[]{
        \includegraphics[width=0.48\textwidth]{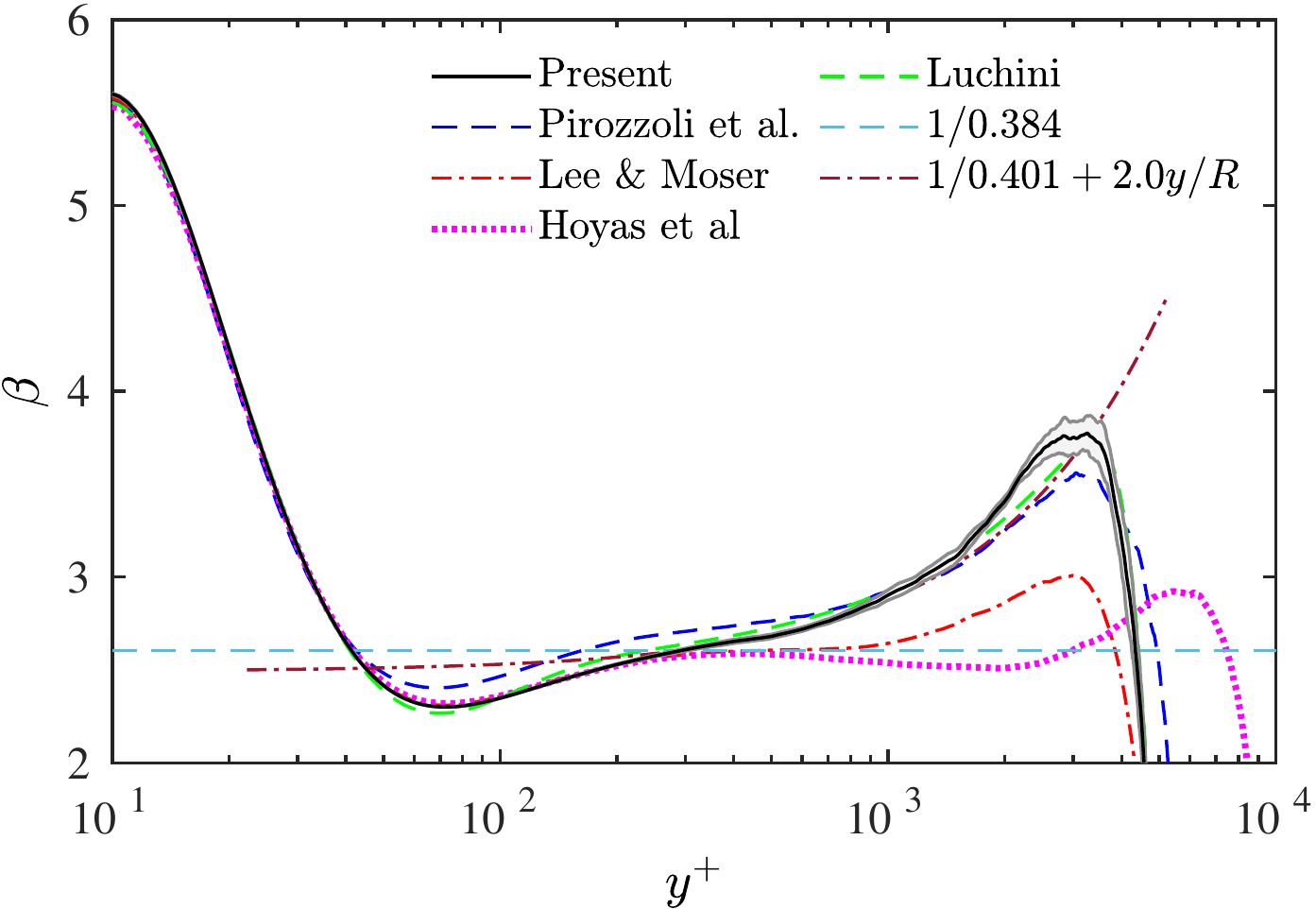}}\\
    \caption{(a) Mean velocity profiles $U^+$  and (b)  log-law diagnostic function $\beta$. Profiles in  (a) are offset in the vertical direction by two wall units, and the  shaded areas in (b) represents the standard deviation of our data at  $Re_\tau=5200$ . }
    \label{fig:umean}
\end{figure}

The $U^+$ profiles at different $Re_\tau$ are  compared to  previous DNS data in figure \ref{fig:umean}(a).
First, as expected,~$U^+$ for pipe flows has a stronger wake when compared with the channel data by \cite{lee2015direct}.
Second, our  data  in the outer region agree well with \cite{el2013direct}'s, but not 
 with  \cite{pirozzoli2021one}.
This discrepancy was also noted by
\cite{pirozzoli2021one}, who found that  their data, along with  those of \cite{wu2008direct} and  \cite{ahn2013direct},  differ from those of \cite{el2013direct} and \cite{chin2014reynolds}.
Again, this disparity seems to be  due to the numerical methods, where  all of the former used low order finite-difference methods, while the latter used  high-order spectral-element methods. 

The log-law diagnostics function $\beta$  is  shown in figure \ref{fig:umean}(b) for all high DNS data (i.e. $Re_\tau>5000$).
Interestingly, $\beta$ for our $Re_\tau=5200$ agrees with the  channel data  by  \cite{lee2015direct} and \cite{hoyas2022wall} up to $y^+\approx250$.
This suggests a near-wall universality of the inner scaled mean velocity -- similar to that observed by \cite{monty2009comparison}.
For all  three cases, the trough is $\beta\approx2.30$, located at  $y^+\approx70$.
However, the data of \cite{pirozzoli2021one} deviates from others for $y^+>40$ and  has a larger magnitude of the trough, which is somehow consistent with the slight upward shift of the $U^+$ observed in Fig \ref{fig:umean}(a). 
This discrepancy is significantly larger than the statistical uncertainty (Appendix \ref{sec:appa}). 
Unlike channel flow, where a plateau starts to develop for $Re_\tau\ge 5200$, there is no plateau for pipe flow -- suggesting that the minimum $Re_\tau$  for $U^+$ to develop a  logarithmic region should be higher  in the pipe than in the channel. 
The~$\beta$ in the wake component of the pipe  is distinctly larger than the channel,  implying a notable difference in flow structures in the core region between these two flows \citep{chin2014reynolds}.

High order corrections to the log-law relation \eqref{eqn:log} were sometimes introduced to better describe  the mean velocity profile in the overlap region \citep{buschmann2003generalized,luchini2017universality,cantwell2019universal}.  
For example, based on refined overlap arguments expressed by \cite{afzal1973analysis},   \cite{jimenez2007we} proposed the following  the indicator function
\begin{equation}\label{eqn:beta2}
    \beta=\Big(\frac{1}{\kappa_\infty}+\frac{\alpha_1}{Re_\tau}\Big)+\alpha_2 \frac{y}{R},
\end{equation}
where $\alpha_1$ and $\alpha_2$ are adjustable constants, and $\kappa_\infty$ is the asymptotic von K\'{a}rman constant.
\eqref{eqn:beta2} allows for a $Re$-dependence of $\kappa=\kappa_\infty+(\alpha_1/Re_\tau)^{-1}$ and introduces a linear dependence on  $y$.
By fitting our $Re_\tau=5200$ data in the region between $y^+=300$ and $y=0.16$, we obtain $\kappa=0.401$ and $\alpha_2=2.0$. 
This $\kappa$  is very close to $0.399$ estimated from the friction factor relation \eqref{eqn:prand}  and   $0.402$ reported by \cite{jimenez2007we} using channel  data of  $Re_\tau=1000$  from \cite{del2004scaling} and $Re_\tau=2000$  from \cite{hoyas2006scaling}. It is slightly larger than $0.387$ by \cite{pirozzoli2021one} and $0.384$ by \cite{lee2015direct}.
In addition,  $\alpha_2$ is generally much larger in the pipe than in channel flow -- suggesting a strong geometry effect on $\beta$.
The value of $\alpha_2=2$  is consistent with the finding  by 
 \cite{luchini2017universality}, who suggested that the logarithmic law of the velocity profile is universal across different geometries of wall turbulence, provided the perturbative effect of the  pressure gradient is taken into consideration.
Furthermore, a  good collapse can be observed between our data and the analytical prediction by  \cite{luchini2017universality}; see, figure \ref{fig:umean}(b).



\subsection{Reynolds stresses}

\begin{figure}
    \centering

    \subfloat[]{
        \includegraphics[width=0.48\textwidth]{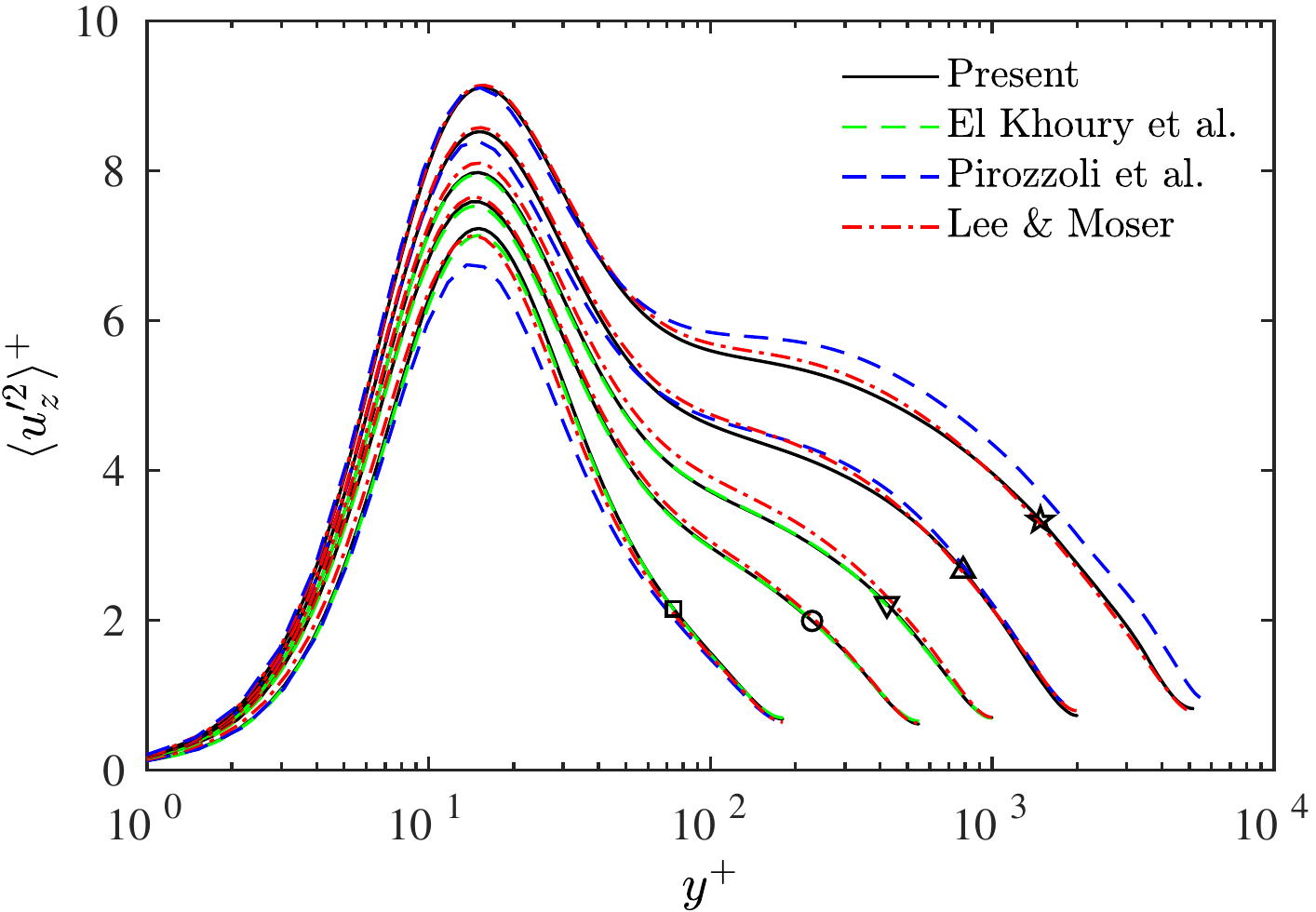}}
     \subfloat[]{
        \includegraphics[width=0.48\textwidth]{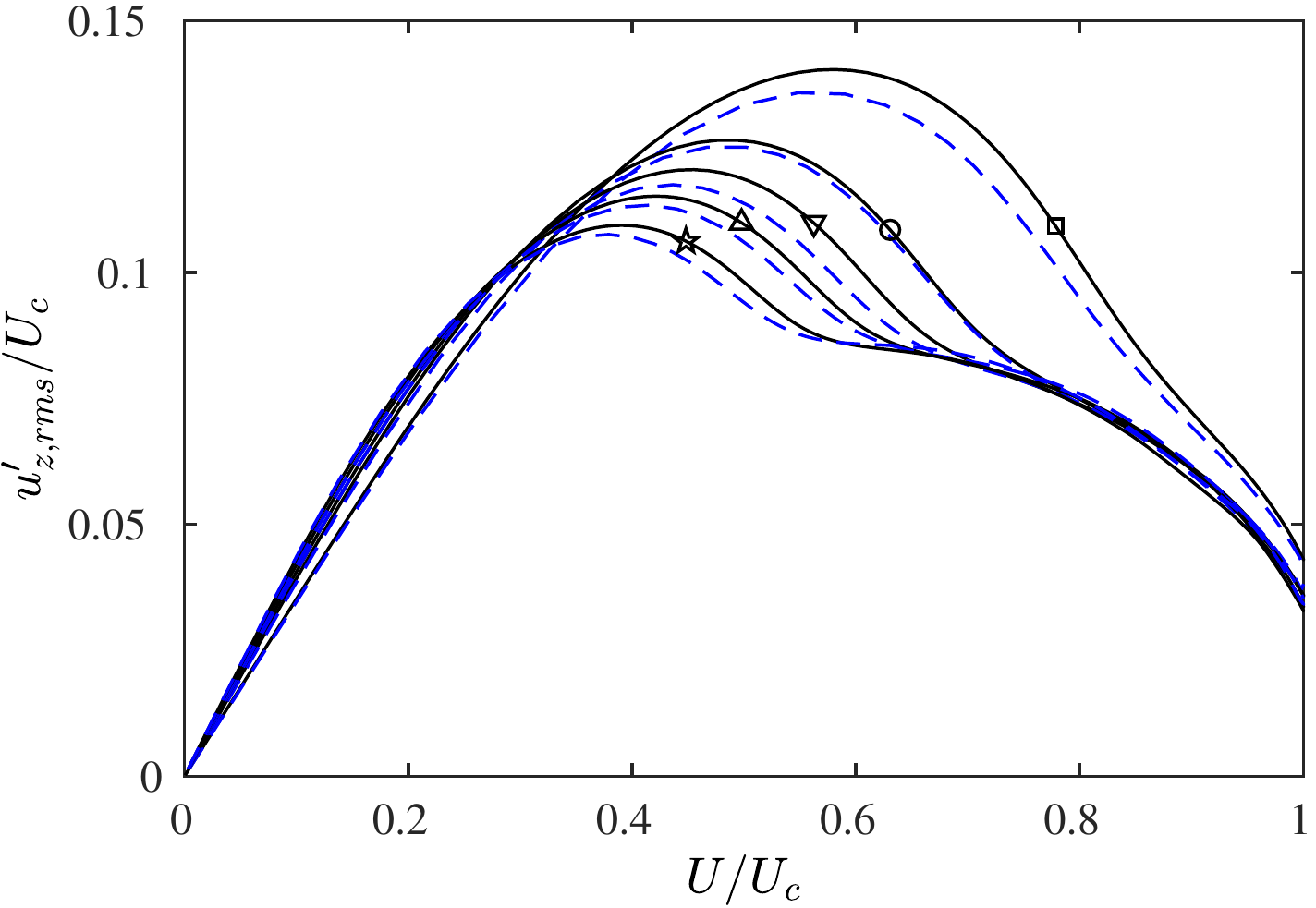}}   \\ 
    \subfloat[]{
        \includegraphics[width=0.6\textwidth]{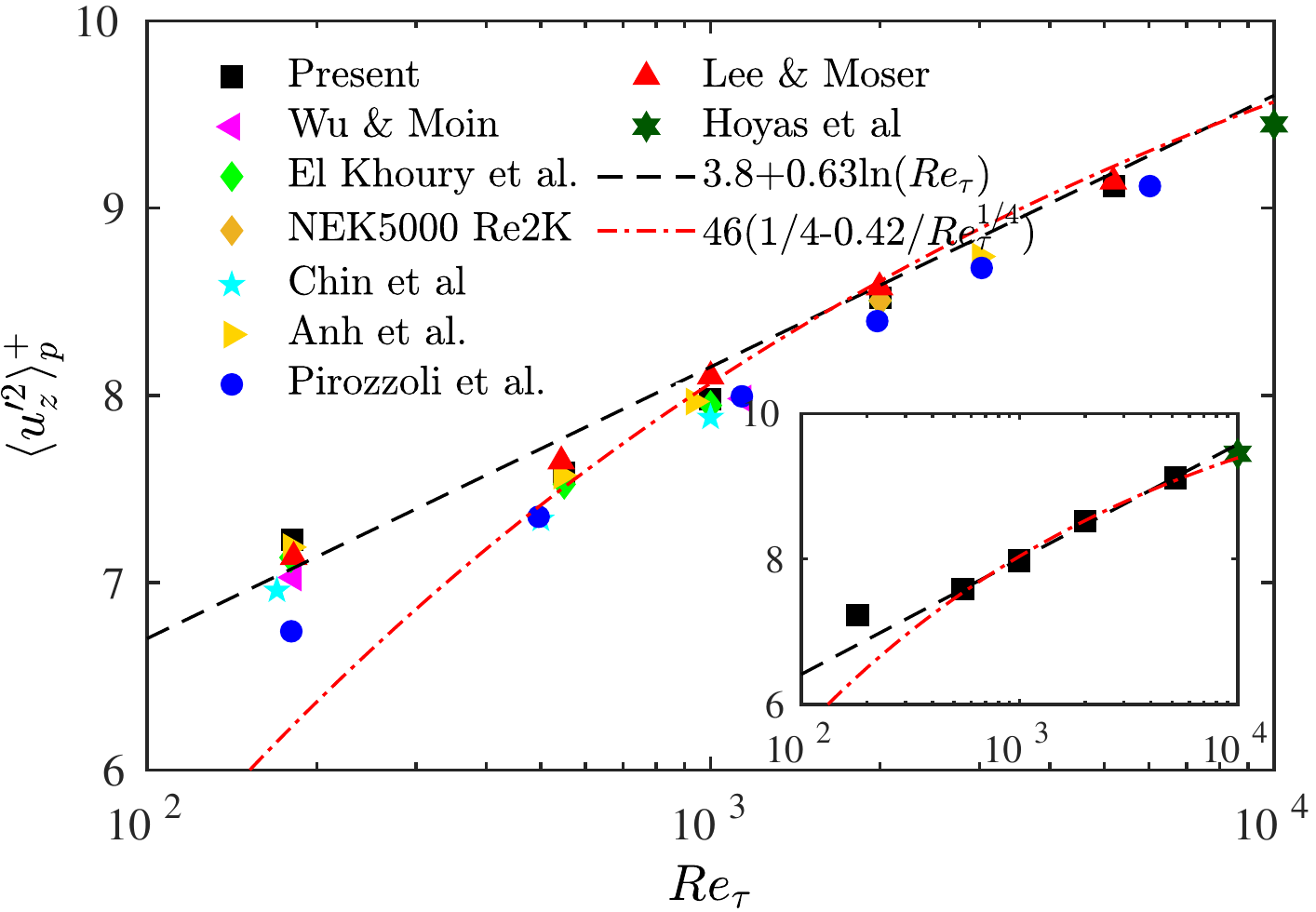}}\\
   
    \caption{(a) Axial velocity variance $\langle u'^2_z  \rangle^+$ as a function of $y^+$; (b) the diagnostic plot depicting $u'_{z,rms}/U_c$ as a function of $U^+/U_c$, and (c) the inner peak of  axial velocity variance $\langle u'^2_z \rangle^+_p$ as a function of $y^+$.  The dashed and dashed-dotted lines in the insert of (c) denote $\braket{u'^2_z}^+_p=3.251+0.687\ln(Re_\tau)$ and $\braket{u'^2_z}^+_p=11.132-17.402Re^{-1/4}_\tau$, respectively. }
    \label{fig:rms}
\end{figure}

\begin{figure}
    \centering
    \subfloat[]{
        \includegraphics[width=0.48\textwidth]{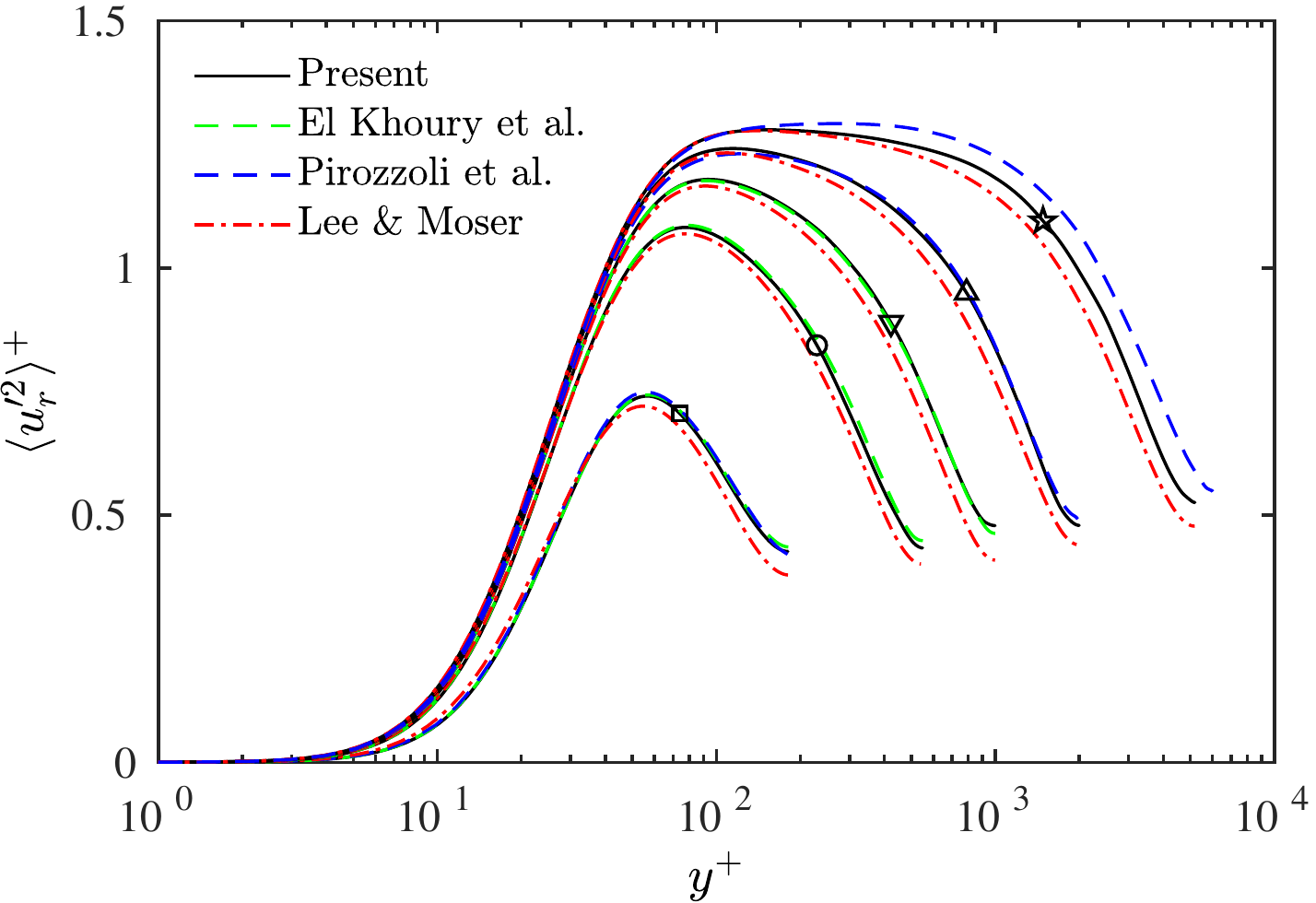}}
    \subfloat[]{
        \includegraphics[width=0.48\textwidth]{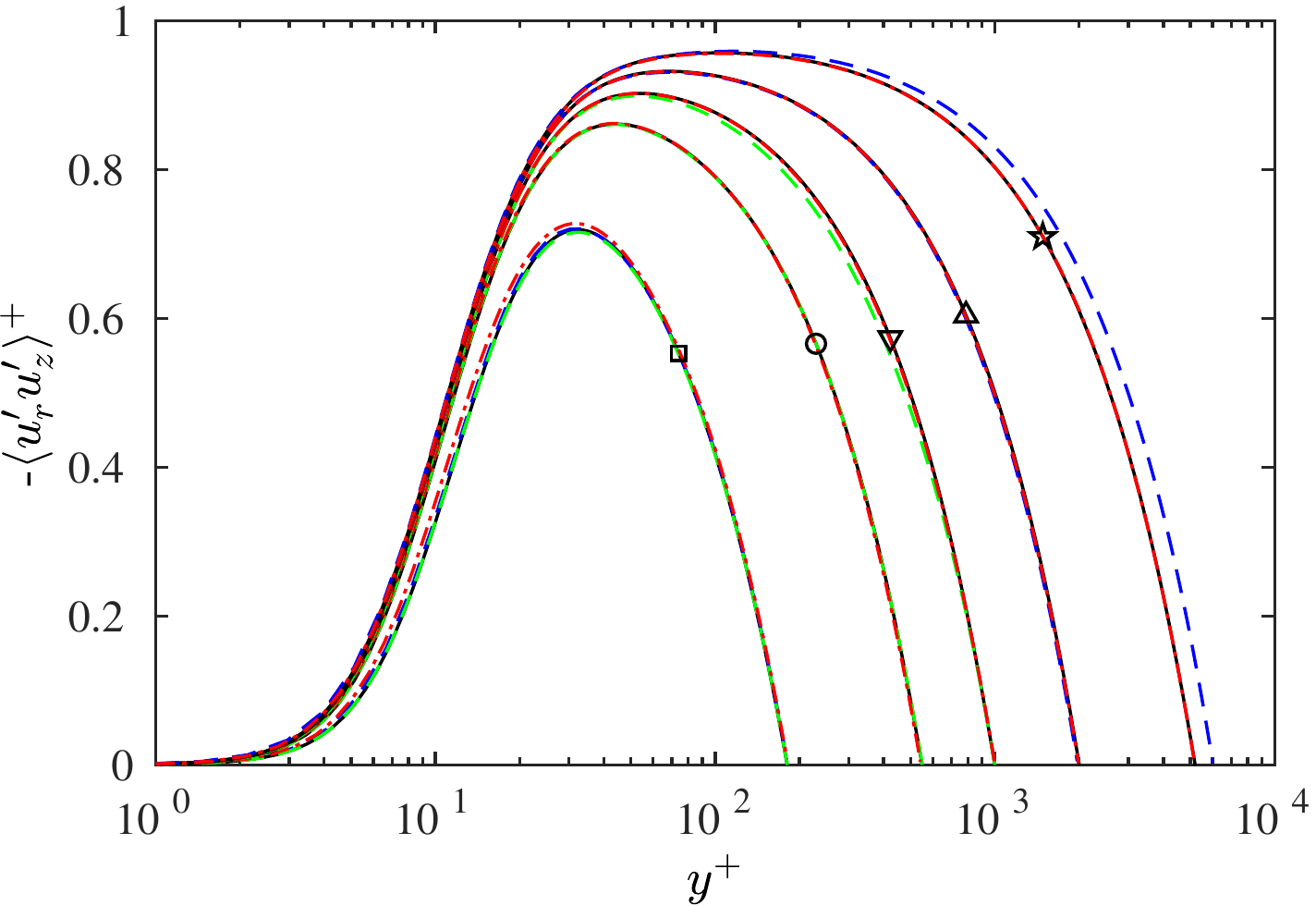}}\\
    \caption{(a) Radial velocity variance $\langle u'^2_r  \rangle^+$ and  (b) Reynolds shear stress $\langle u'_r u'_z \rangle^+$ as a function of $y^+$. }
    \label{fig:uvrms}
\end{figure}

\begin{figure}
    \centering

    \subfloat[]{
        \includegraphics[width=0.48\textwidth]{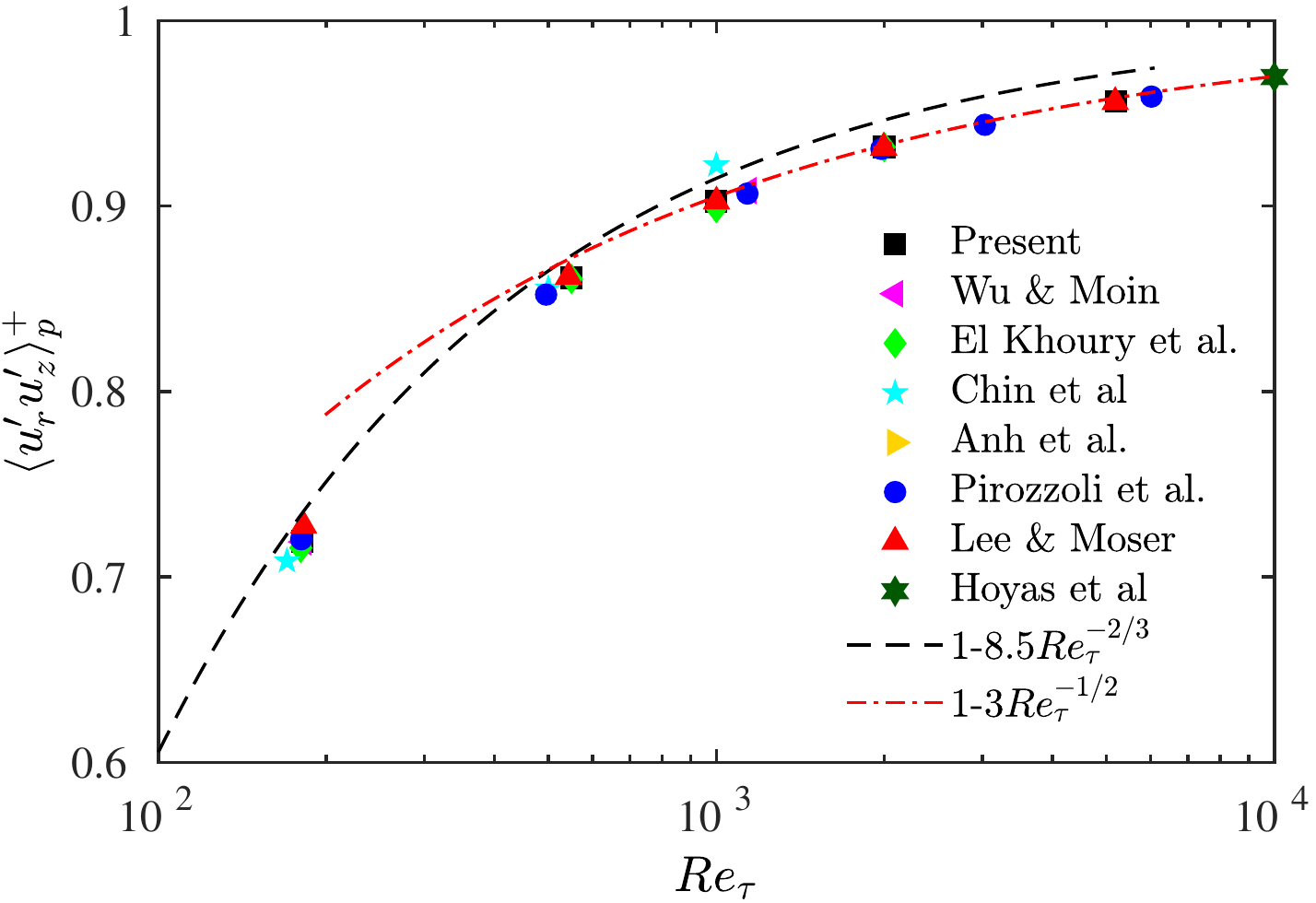}}
    \subfloat[]{
        \includegraphics[width=0.48\textwidth]{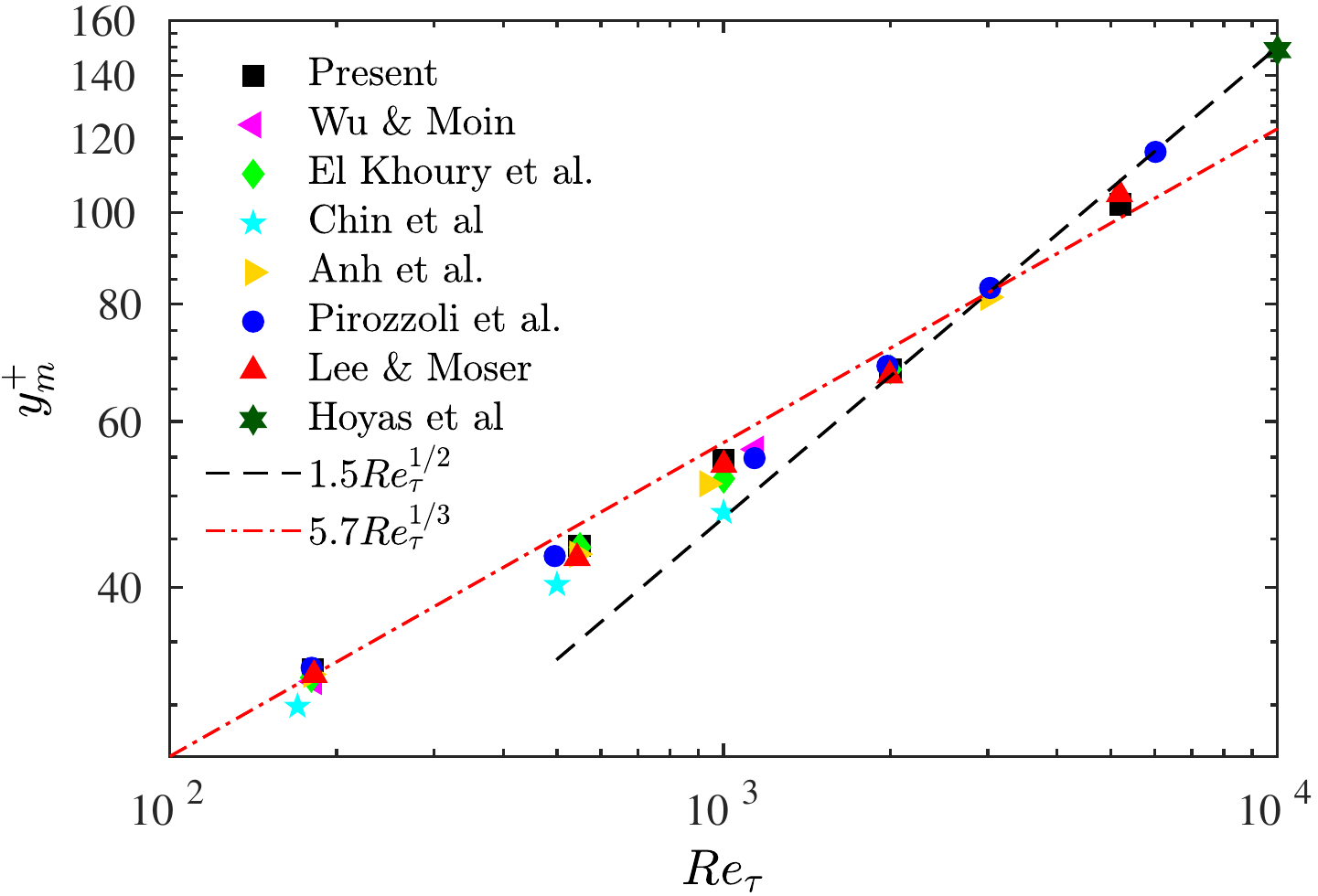}}\\
    \caption{Reynolds number dependence of (a) the peak of the Reynolds shear stress $\langle u'_r u'_z \rangle^+$  and (b) the corresponding peak location in wall units. }
    \label{fig:uvpeak}
\end{figure}

\begin{figure}
    \centering

        \subfloat[]{
        \includegraphics[width=0.48\textwidth]{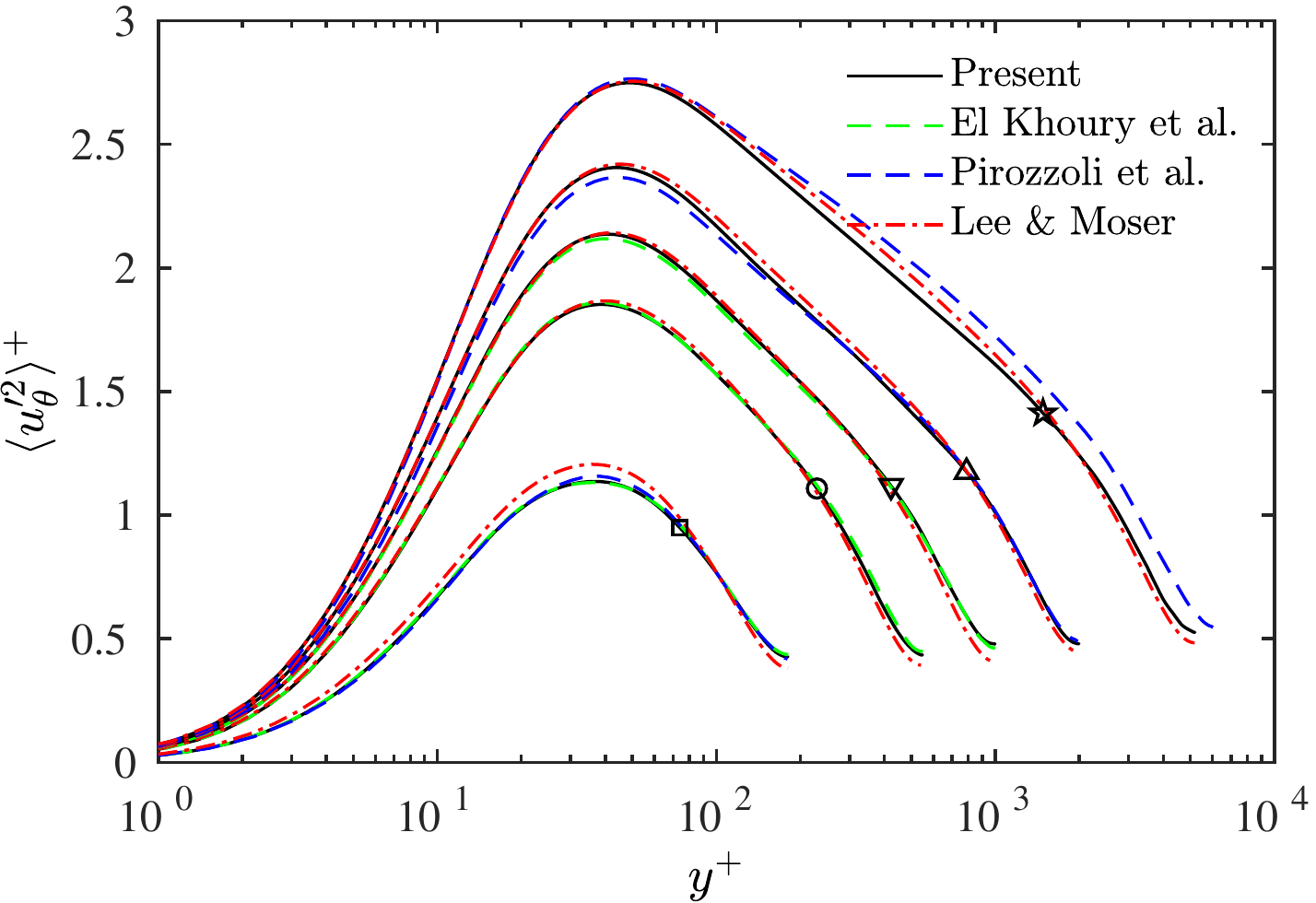}}
    \subfloat[]{
        \includegraphics[width=0.48\textwidth]{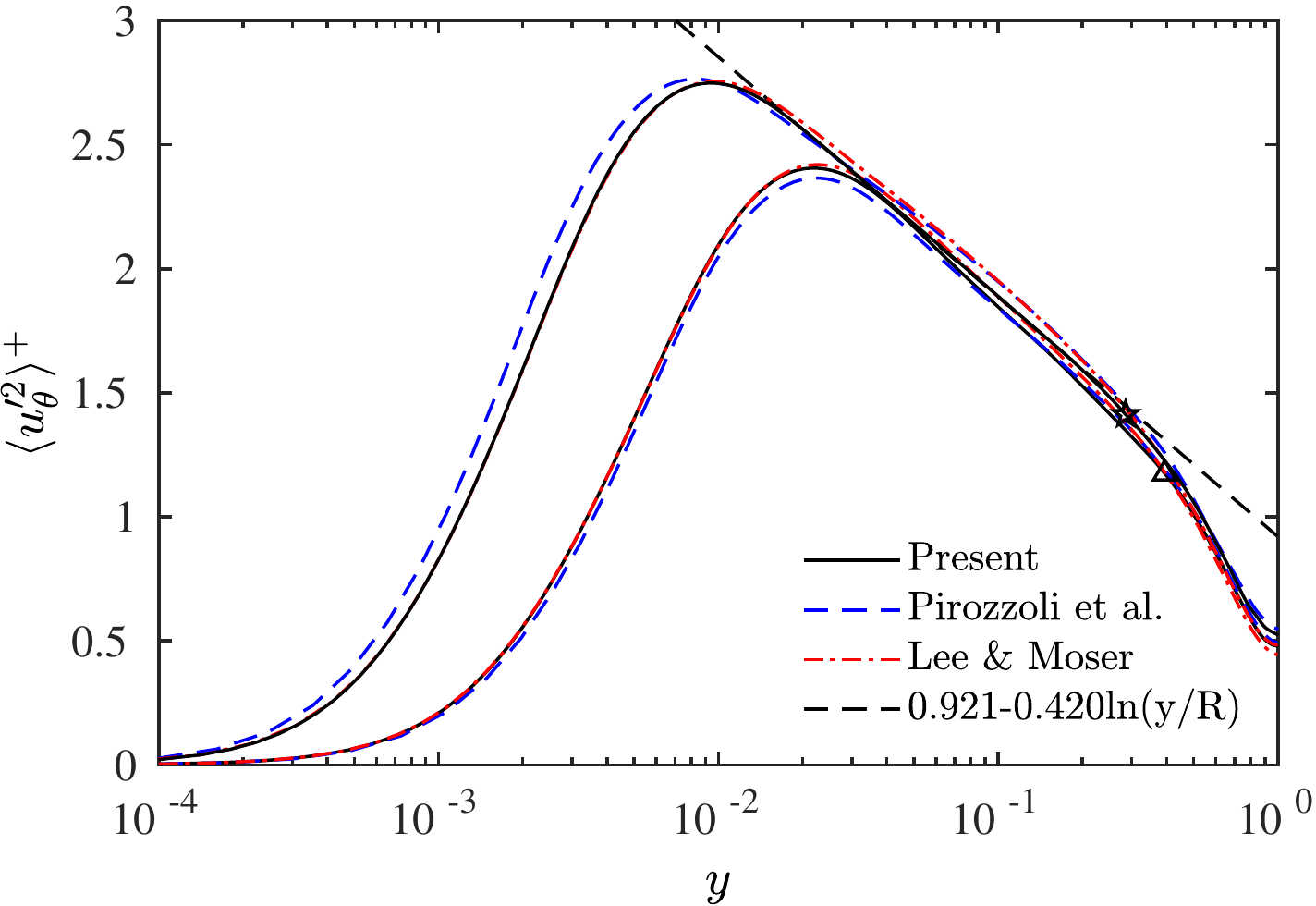}}\\
    \caption{Azimuthal velocity variance $\langle u'^2_\theta  \rangle^+$ as a function of (a) $y^+$ and (b) $y/R$. Note that in (b) only the two highest $Re_\tau$ cases are included. }
    \label{fig:wrms}
\end{figure}

The non-zero components of the Reynolds stress tensor (or the velocity variances and covariance) are examined in this section (figures  \ref{fig:rms}--\ref{fig:wrms}).
For all datasets, the  inner-scaled  velocity variances and covariance increase with $Re_\tau$ in the whole wall-normal range.
In terms of  the axial velocity variance $\braket{u'^2_z}^+$ (figure \ref{fig:rms}a),  our data agrees well with  \cite{el2013direct} but differs from  \cite{pirozzoli2021one}, which is notably smaller in the near wall region, particularly at low $Re_\tau$. 
For the highest $Re_\tau$ cases, the agreement is reasonably good near the wall. However, note that  \cite{pirozzoli2021one}'s simulation is  at a slightly higher $Re_\tau$ (i.e. $\sim6000$).
The differences between our and  \cite{pirozzoli2021one}'s cases can be better highlighted in the diagnostic  plot (figure \ref{fig:rms}b), where the root mean square (r.m.s.)  axial velocity fluctuation $u'_{z,rms}$ is plotted against the mean velocity $U^+$.
The diagnostic plot was introduced by \cite{alfredsson2010diagnostic} as a means to assess if the  mean velocity and velocity  fluctuation profiles behave correctly without the need to determine the friction velocity or the wall position. 
Consistent with the observation in  \cite{alfredsson2010diagnostic}, the diagnostic plot collapses in the outer parts  for $Re_\tau>180$, and has a clear $Re$ trend around the peak value.
Most importantly, the data by \cite{pirozzoli2021one} is consistently lower than ours, particularly in the near-wall region. 
Such inconsistency is also observed for $\braket{u'^2_\theta}^+$ (figure  \ref{fig:wrms}). 
The agreement for $\braket{u'^2_r}^+$ (figure \ref{fig:uvrms}a) is reasonably good among different pipe flow datasets, which is slightly larger than the channel, particularly in the outer region. 
The Reynolds shear stress $\braket{u'_r u'_z}^+$  shows the best agreement among different datasets, even  including the channel.

Let us focus now on the inner peak of the axial velocity variance $\braket{u'^2_z}^+$.
The inner peak is assumed to  increase logarithmically with $Re_\tau$ -- similar as the wall shear stress fluctuations due to the increased modulation effect of the large-scale structures  in the logarithmic layer \citep{marusic2019attached}.
\cite{chen2021reynolds} recently suggested that the growth of $\braket{u'^2_z}^+$ would eventually saturate.
The argument is based on the balance between the viscous diffusion and dissipation at the wall and  the Taylor series expansion of the axial velocity variance near the wall, given as 
\begin{eqnarray}\label{eqn:uzz}
\braket{u'^2_z}^+\sim D^+_{z,w}y^{+2}= \epsilon^+_{z,w} y^{+2}.
\end{eqnarray}
Note that a similar expression is obtained in \cite{smits2021reynolds}, where the axial wall dissipation is used instead, i.e. $\braket{u'^2_z}^+\sim \langle \tau'^2_{z,w} \rangle^+y^{+2}$. 
If the  assumption of the boundness of wall dissipation (\ref{eqn:dissp}) is valid and the  inner peak location of $\braket{u'^2_z}^+$ (denoted as $y^+_{z,p}$) is  independent of $Re_\tau$, then \eqref{eqn:uzz} suggests that the peak of axial velocity variance should also be bounded in a similar defect power form as the wall shear stress fluctuation:
\begin{eqnarray}\label{eqn:upower}
\braket{u'^2_z}^+_p=M- NRe_\tau^{-1/4},
\end{eqnarray}
where  $M$ is the asymptotic value and $N$ is the coefficient. 

This validity of  \eqref{eqn:upower} was recently challenged by \cite{pirozzoli2021one}, who, based on their data,  observed a  slight increase of  $y^+_{z,p}$ with $Re_\tau$, from $y^+_{z,p}=14.28$ for $Re_\tau\approx500$ to $15.14$ for $Re_\tau\approx 6000$.
We emphasize that such variation of $y^+_{z,p}$ with $Re_\tau$ is not observed in our case, where much finer near-wall resolutions are  used than in  \cite{pirozzoli2021one}. 
The value of  $y^+_{z,p}$ is approximately $15$ for all $Re_\tau$ (e.g. $y^+_{z,p}=15.07$, $15.03$, $15.50$ for $Re_\tau=180$, $2000$, and $5000$, respectively)  -- akin to the findings  by many others  \citep{moser1999direct,jimenez2010turbulent,chin2014reynolds, smits2021reynolds}).

Figure \ref{fig:rms}(c) shows the $\braket{u'^2_z}^+_p$ for all the DNS data listed in table \ref{tbl:Num2}, along with the logarithmic law $\braket{u'^2_z}^+_p=3.8+0.64\ln(Re_\tau)$  by \cite{marusic2017PRF} and the power law $\braket{u'^2_z}^+_p=11.5-19.32Re_\tau^{-1/4}$ by \cite{chen2021reynolds}. The difference between different DNS datasets is relatively small, except for those from  \cite{pirozzoli2021one}, which are consistently lower than others for all $Re_\tau$. Note that such discrepancy is much larger than the uncertainty (standard deviation), which is less than $0.5\%$ (see table \ref{tbl:Num3}).  
Both the logarithmic and defect power laws  fit well with the data  at the high $Re_\tau$ range but have certain discrepancies at the low $Re_\tau$. 
It  suggests that there might exist a transitional scaling   -- similar to that found for the Reynolds shear stress \citep{chen2019non}. 
The parameters in these two scaling laws can be adjusted to better fit our  dataset. 
The inset shows the fitting results for our data without the one at $Re_\tau=180$: $\braket{u'^2_z}^+_p=3.251+0.687\ln(Re_\tau)$ and $\braket{u'^2_z}^+_p=11.132-17.402Re^{-1/4}_\tau$.
With these new constants, the agreement is improved for both  scaling laws. 
In summary, for the $Re_\tau$ range studied, both  scaling laws can provide good match with $\braket{u'^2_z}^+_p$ data when the fitting parameters are properly adjusted. 
Data at even higher $Re_\tau$ is required to determine which law is more consistent with the data.

According to Townsend's attached eddy hypothesis, at sufficiently high $Re$, the Reynolds stress  components in a certain $y$ range   satisfy
\begin{eqnarray}
\braket{u'^2_z}^+&=&A_1-B_1 \ln(y/R),\\
\braket{u'^2_z}^+&=&A_2,\\
\braket{u'^2_\theta}^+&=&A_3-B_3 \ln(y/R),\\
\braket{u'_r u'_z}^+&=&-1,
\end{eqnarray}
where $A_i$ and $B_i$ are universal constants.

Consistent with these relations, the radial velocity variance $\braket{u'^2_r}^+$ slowly develops a flat region as $Re_\tau$ increases. 
In addition, the Reynolds shear stress $\braket{-u'_ru'_z}^+$ profiles also tend to become flattened at higher $Re_\tau$. 
As noted by \cite{afzal1982fully}, the peak  Reynolds shear stress at high $Re_\tau$ follows $\braket{-u'_ru'_z}^+_p\approx 1-2/\sqrt{\kappa Re_\tau}$, and the corresponding  position  $y^+_m$ shifts away from the wall following  $y^+_m\sim\sqrt{Re_\tau/\kappa}$.
\cite{chen2019non} suggested that there is a non-universal scaling transition, where the peaks at low $Re_\tau$ scales as $\braket{u'_ru'_z}^+_p\approx  Re_\tau^{-2/3}$ and their locations scales as $y^+_m \sim Re_\tau^{1/3}$.
Figure \ref{fig:uvpeak} shows   $\braket{-u'_ru'_z}^+_p$ and the corresponding  $y^+_m$ as a function of $Re_\tau$.
For $Re_\tau>1000$, the $Re^{-1/2}$ for $\braket{-u'_ru'_z}^+_p$ and $Re^{-1/2}$ for $y^+_m$ are satisfied with good accuracy, and at the low $Re_\tau$ range, the $Re^{-2/3}_\tau$ for $\braket{-u'_ru'_z}^+_p$ and $Re^{-1/3}_\tau$ for $y^+_m$ scalings proposed by \cite{chen2019non}  also yield a good agreement.

Regarding the axial velocity variance $\langle u'^2_z \rangle^+$, no clear logarithmic region is observed  for the $Re_\tau$ range considered here.
As discussed in  \cite{lee2015direct}, $Re_\tau=5200$ is not quite high enough to exhibit such a region. Based on the Superpipe data, \cite{marusic2013logarithmic} suggested that a sensible logarithmic layer emerges only for $Re_\tau>10^4$.
Consistent with the findings in  \cite{lee2015direct} and \cite{pirozzoli2021one}, the azimuthal velocity variance $\langle u'^2_\theta  \rangle^+$ (figure \ref{fig:wrms}b) develops the logarithmic layers even at lower $Re_\tau$. Fitting the $Re_\tau=5200$ data in the range between $0.02\le y \le 0.2$ yields $A_3=0.921$, $B_3=0.420$, which is close to $A_3=1.0$, $B_3=0.40$ by \cite{pirozzoli2021one}, and   between $A_3=1.08$, $B_3=0.387$ obtained by \cite{lee2015direct} for the turbulent channel and $A_3=0.8$, $B_3=-0.45$ by \cite{sillero2013one} in the turbulent boundary layer.

\subsection{Production and dissipation of the turbulent kinetic energy}

\begin{figure}
    \centering
    \subfloat[]{
        \includegraphics[width=0.48\textwidth]{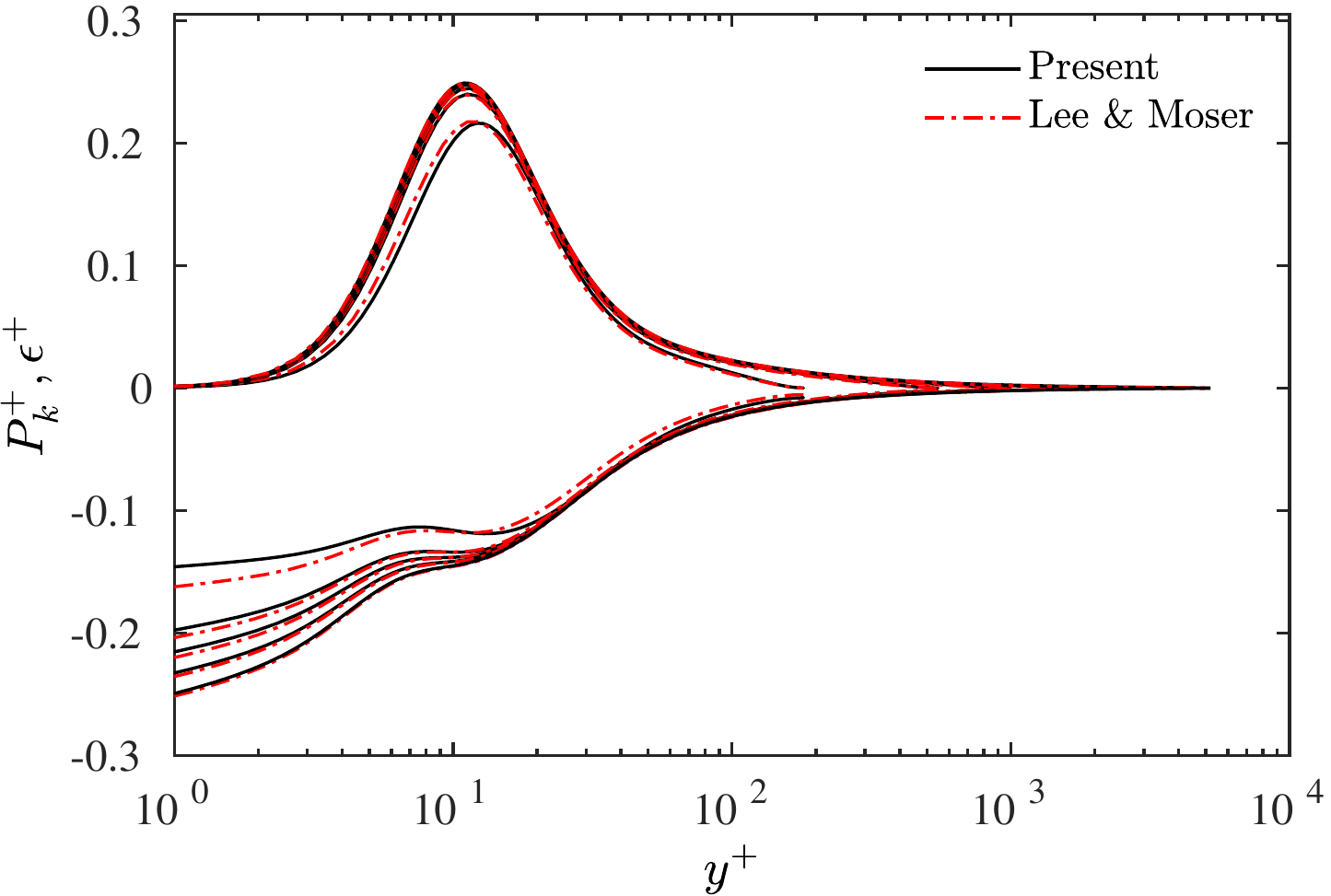}}
    \subfloat[]{
        \includegraphics[width=0.48\textwidth]{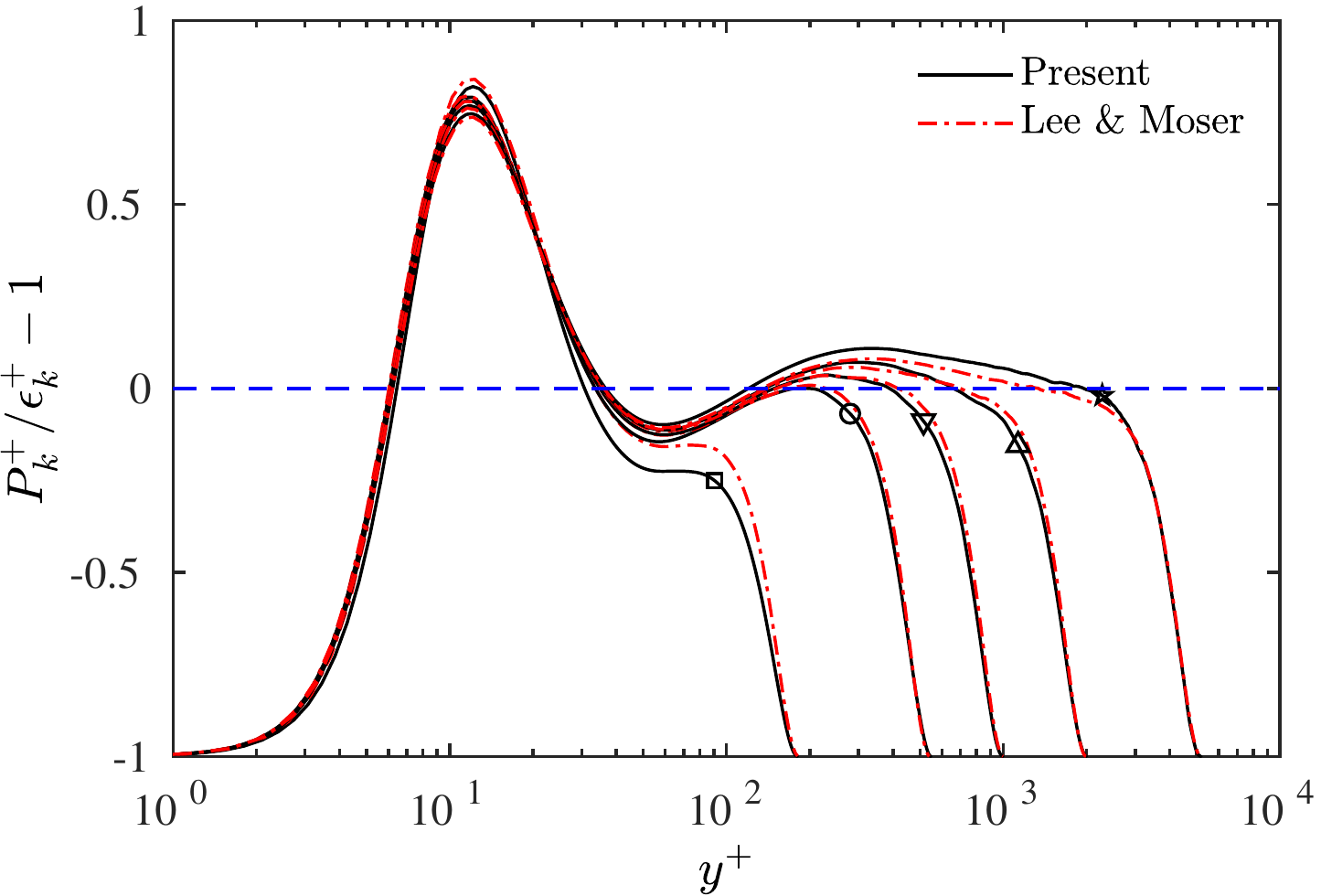}}\\
    \caption{(a) Production $P^+_k$ and dissipation $\epsilon^+_k$ of the turbulent kinetic energy, and (b) balance of production and dissipation $P^+_k/\epsilon^+_k-1$ as a function of $y^+$. }
    \label{fig:prod}
\end{figure}

Figure \ref{fig:prod}(a) shows the production $P^+_k$ and dissipation $\epsilon^+_k$ of the turbulent kinetic energy (i.e. $k^+=(\langle u'^2_r \rangle^++\langle u'^2_\theta \rangle^++\langle u'^2_z  \rangle^+)/2$). Other terms in the transport equations of the turbulent kinetic energy and individual Reynolds stress components are available at https://dataverse.tdl.org/dataverse/turbpipe.
The production $P^+_k$ has a peak at around $y^+\approx11$, and the corresponding magnitude approaches the asymptotic value $1/4$ as $Re_\tau$
increases.
Despite notable differences in the mean axial/streamwise velocity profile   observed in the outer region, $P^+_k$ is quite similar between  pipes and channels. 
It explains why the higher velocity gradient of the pipe
does not contribute an effect to the turbulence intensities.
The magnitude of dissipation $\epsilon^+_k$ continuously increases with $Re_\tau$, and the difference between  the pipe and channel is mainly in the near wall region and decreases with increasing $Re_\tau$.

At sufficiently high $Re_\tau$, there is an intermediate region where the production balances the dissipation. Recent experimental \citep{Hultmark2012PRL}  and numerical \citep{lee2015direct,pirozzoli2021one} results suggest that such equilibrium between production and dissipation is violated due to the presence of  LSMs and VLSMs.
Figure \ref{fig:prod}(b) shows the relative excess of production over dissipation ($P^+_k/\epsilon^+_k-1$). 
First, there is a  near-wall region ($8\le y^+\le35$), where  $P^+_k$ distinctly exceeds the $\epsilon^+_k$.
At high $Re_\tau$, another region of $P^+_k/\epsilon^+_k>1$ develops, and the  magnitude increases with  $Re_\tau$.
For the $Re_\tau=5200$ case, the peak imbalance is about $11\%$ (located at $y^+\approx330$), which is slightly larger than in the channel (i.e. $8\%)$

\subsection{Mean pressure and rms of pressure fluctuation}

\begin{figure}
    \centering
    \subfloat[]{
        \includegraphics[width=0.48\textwidth]{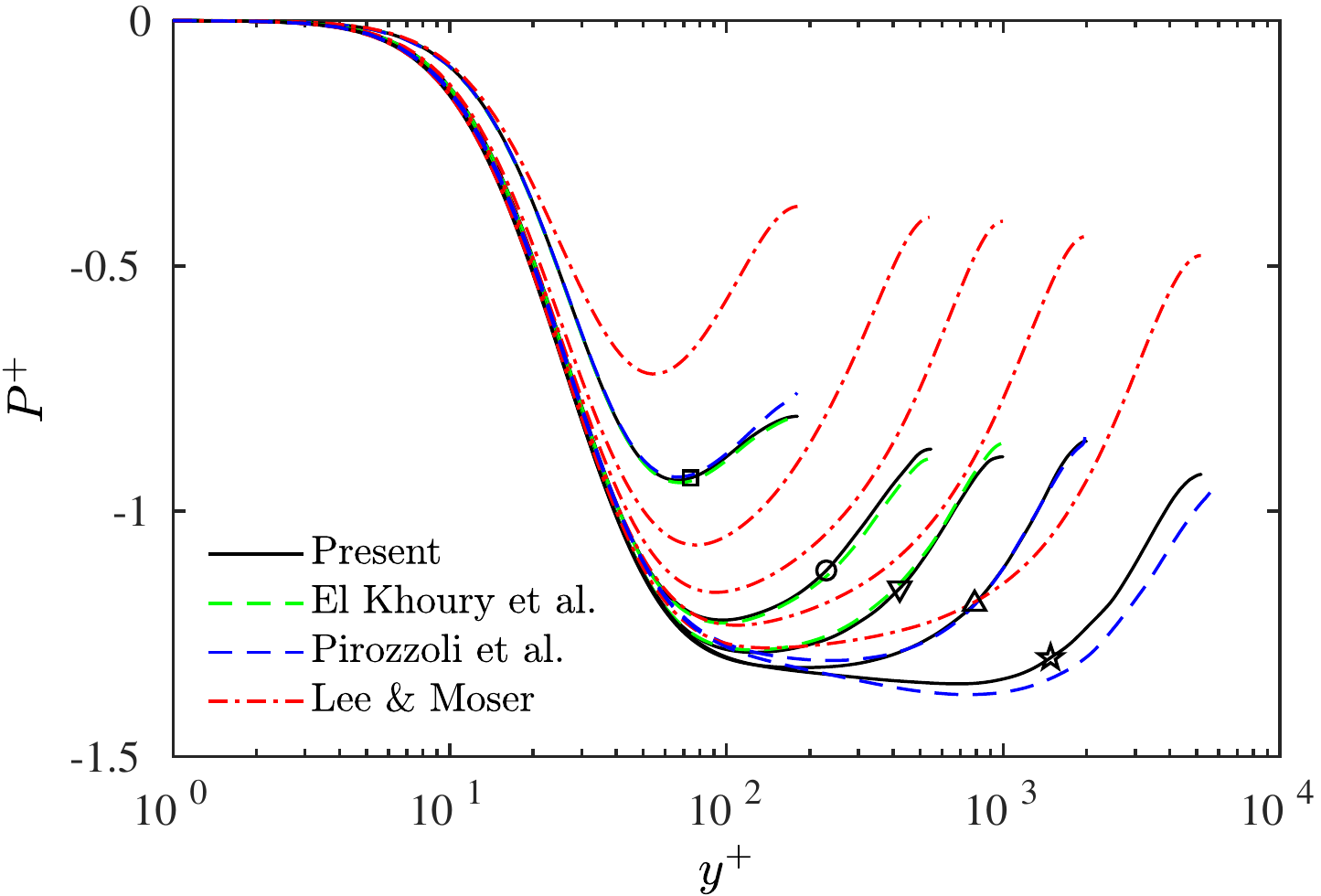}}
    \subfloat[]{
        \includegraphics[width=0.48\textwidth]{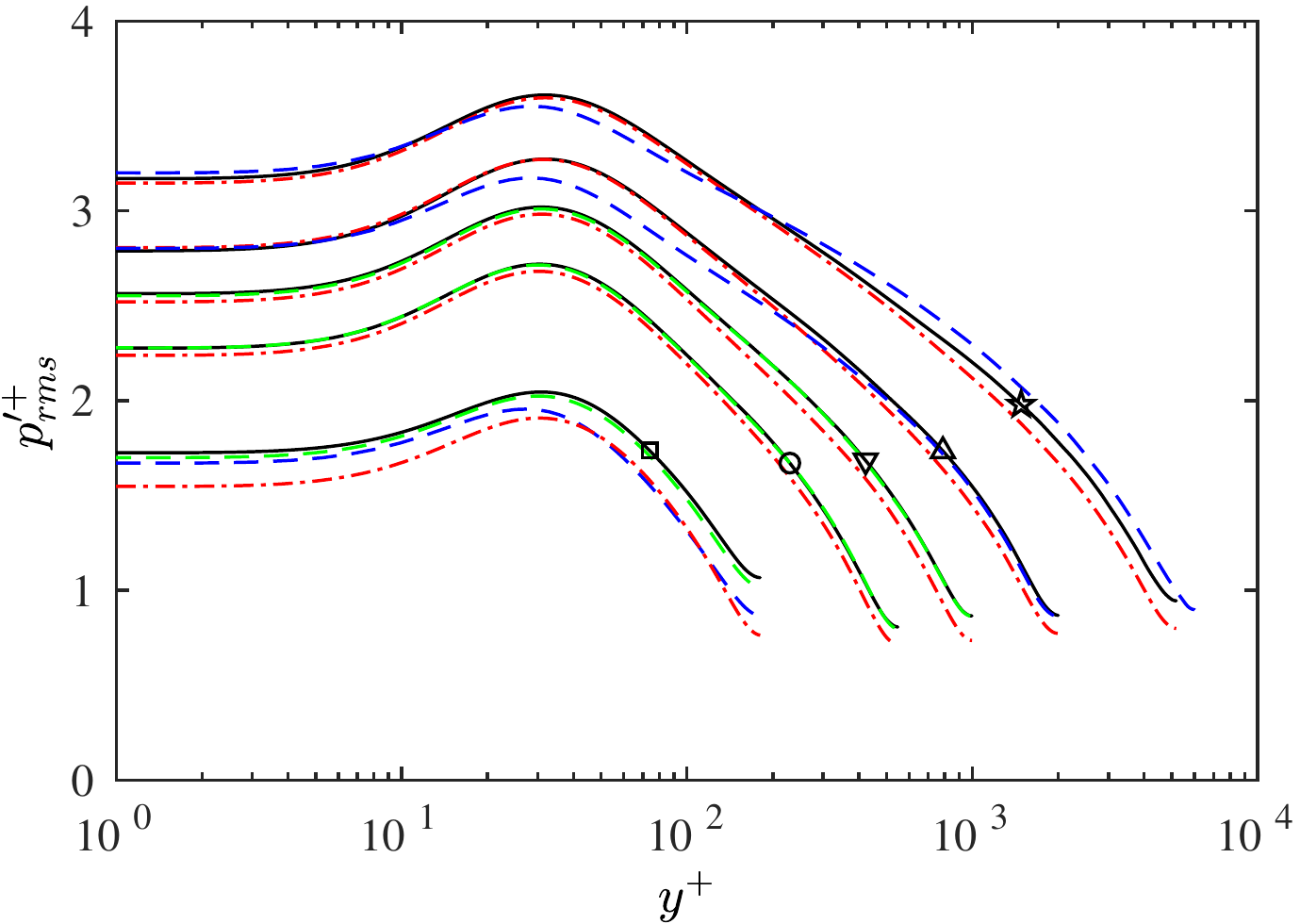}}\\
    \caption{(a) Mean pressure ($P^+$) and (b) r.m.s.  pressure fluctuation ($p^+_{rms}$) as a function of $y^+$. }
    \label{fig:MP}
\end{figure}

The mean pressure  and  r.m.s.  pressure fluctuations are displayed in figure \ref{fig:MP}. 
First, the mean pressure $P^+$ has  different behavior in the outer region between pipe and channel flows, with $P^+$ being substantially lower in the wake of the pipe. 
As discussed in  \cite{el2013direct}, this difference is related to  the mean radial momentum equation, which, in pipe flow, is given as  \citep{hinze1975turbulence}
\begin{eqnarray}\label{eqn:mrp}
\frac{1}{\rho}\frac{\partial P}{\partial r}+\frac{\mathrm{d}}{\mathrm{d} r}\braket{u'^2_r }+\frac{\braket{u'^2_r }-\braket{u'^2_\theta }}{r}=0.
\end{eqnarray}
By changing variables (i.e. $r=R-y$) and then integrating the above equation, the mean pressure for pipe flow with the wall value set to zero can be expressed as 
\begin{eqnarray}\label{eqn:mrp2}
P^+(y)=-\braket{u'^2_r}^++\int^y_0\frac{\braket{u'^2_r }^+-\braket{u'^2_\theta }}{R-y'}^+ \mathrm{d}y'.
\end{eqnarray}
In channel flow, the last term on the left-hand side of  \eqref{eqn:mrp} is absent, and the mean pressure is solely balanced by the wall-normal velocity fluctuation, i.e. $P^+(y)=-\braket{v'^2}^+$. 
From figure \ref{fig:uvrms}(a), it is clear that the wall-normal velocity fluctuation is comparable between pipe and channel flows. However, as $\braket{u'^2_r }^+<\braket{u'^2_\theta }^+$ in pipe flow, the extra term in  \eqref{eqn:mrp2} is zero at the wall and decreases with increasing $y$ --  resulting in a lower pressure in pipes than in channels.

\begin{figure}
    \centering

    \subfloat[]{
        \includegraphics[width=0.48\textwidth]{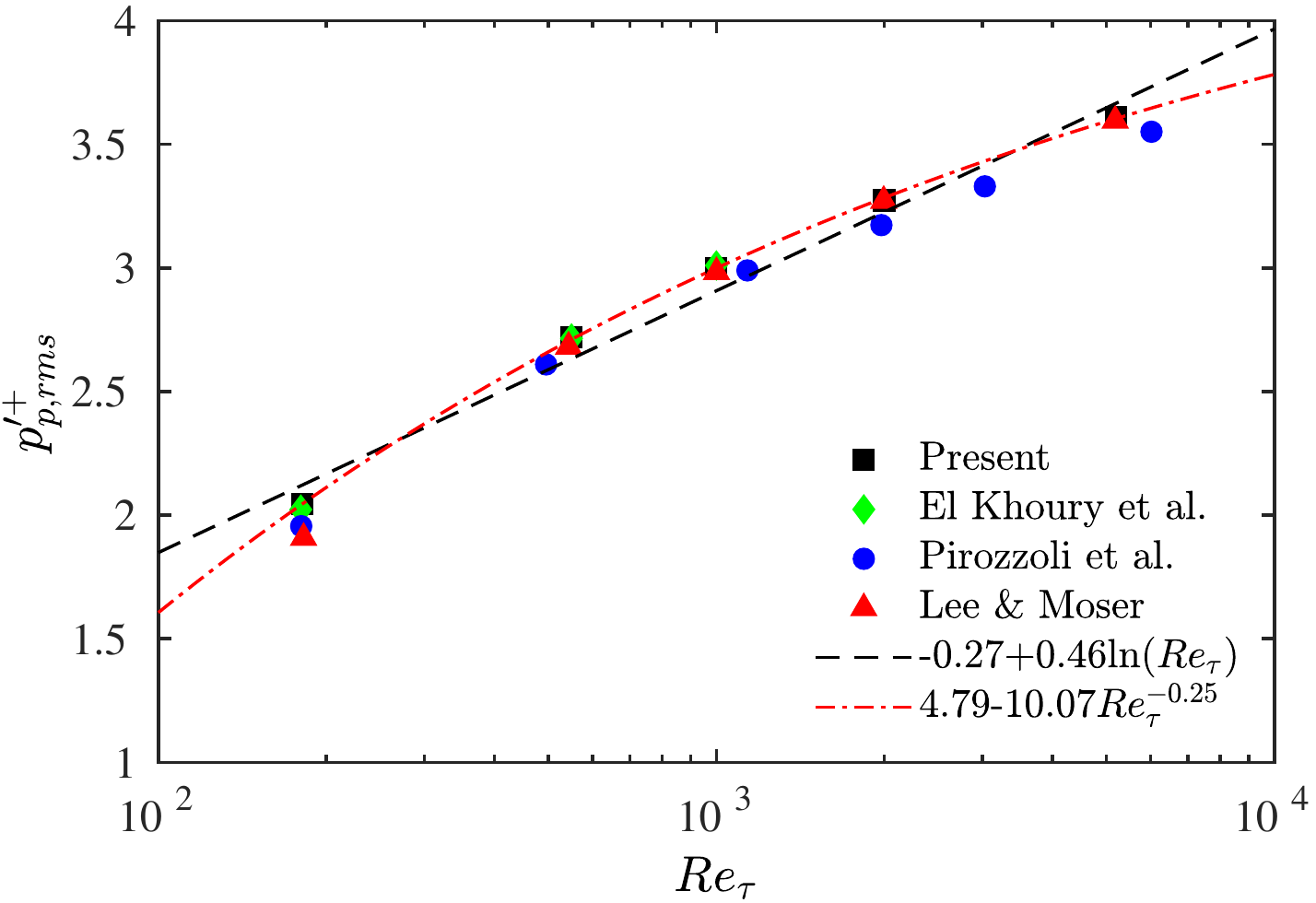}}
    \subfloat[]{
        \includegraphics[width=0.48\textwidth]{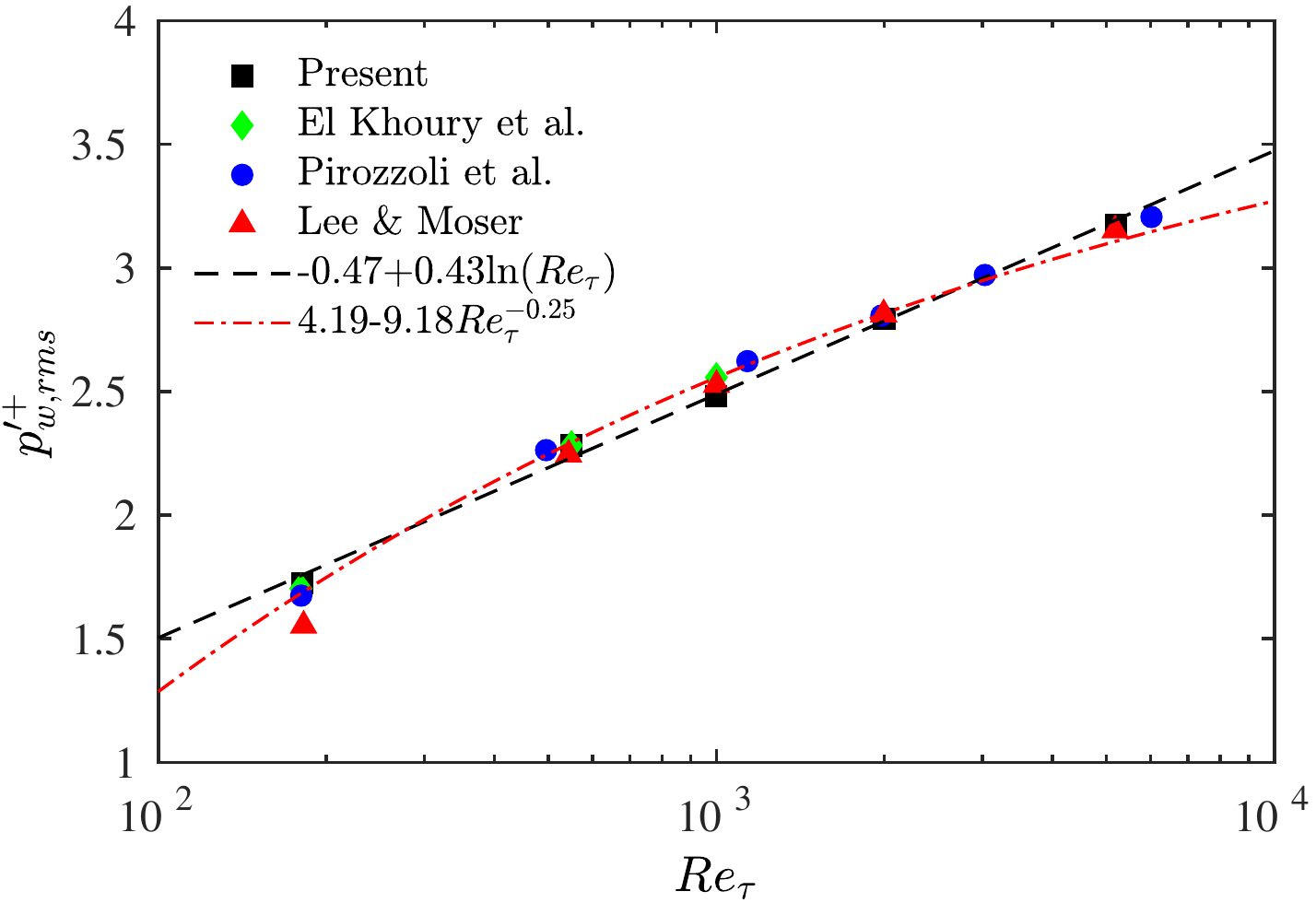}}\\
    \caption{Reynolds number dependence of (a) peak  ($p^+_{p,rms}$)  and (b) wall  ($p'^+_{w,rms}$) values of r.m.s.  pressure fluctuations. }
    \label{fig:Peak}
\end{figure}

Similar to that observed by \cite{el2013direct}, the r.m.s.  pressure fluctuation $p'^{+}_{rms}$  exhibits similar behavior between  the pipe and  channel flows, except for  slightly lower values for the latter. 
Minor differences  are observed between our data and those of \cite{pirozzoli2021one}, particularly near the peak value. 
The difference between our data with the channel data by  \cite{lee2015direct} in the near-wall region decreases with increasing $Re_\tau$. 
Figure \ref{fig:Peak} further shows the peak and wall values of $p'^+_{rms}$, which has similar $Re$-dependence as for other measures, such as wall shear stress fluctuations and axial velocity variances. 
Again, for the $Re_\tau$ studied,  both the logarithmic and defect power laws  fit the data well.

\subsection{Energy spectra}

\begin{figure}
    \centering

    \subfloat[]{
        \includegraphics[width=0.48\textwidth]{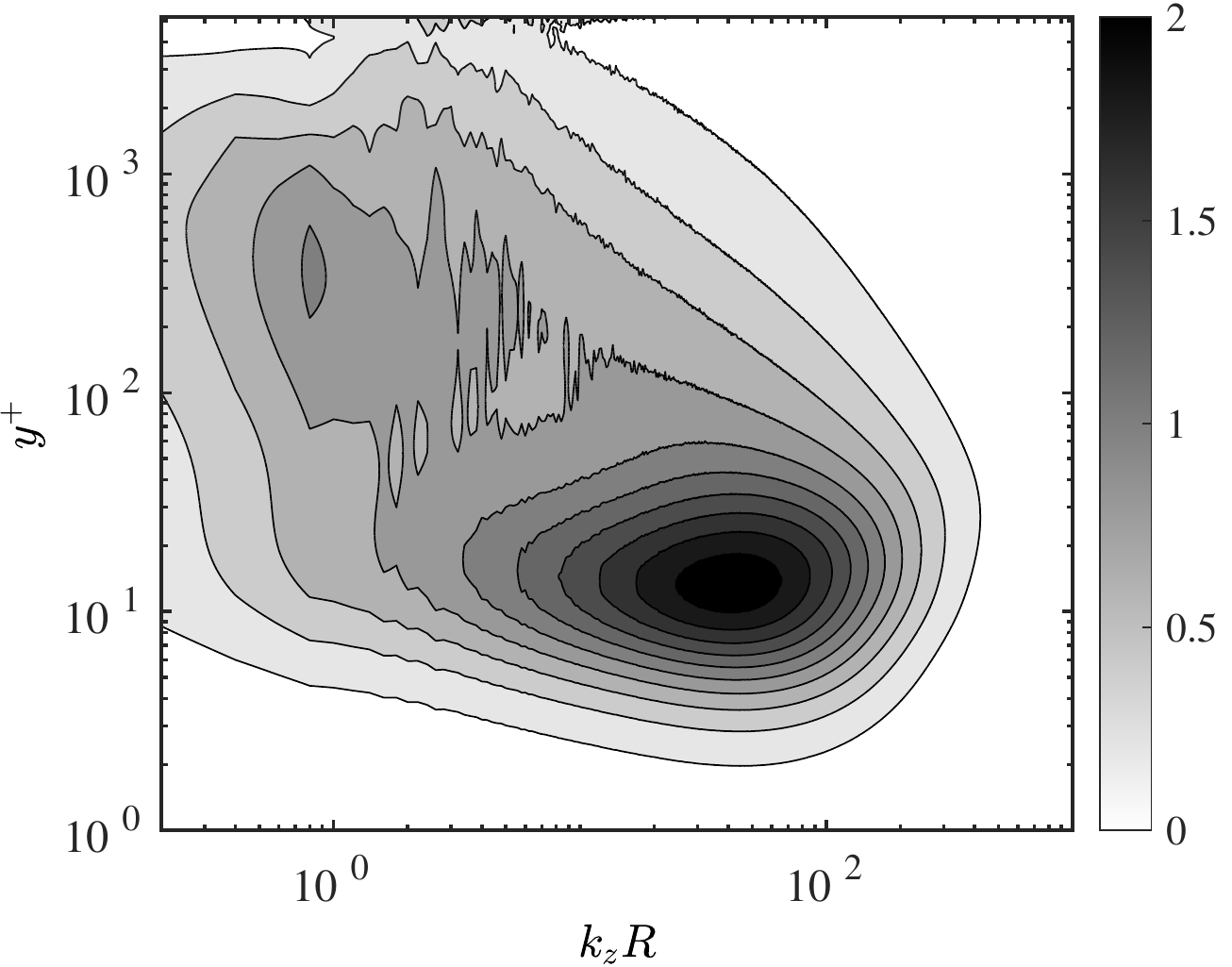}}
    \subfloat[]{
        \includegraphics[width=0.5\textwidth]{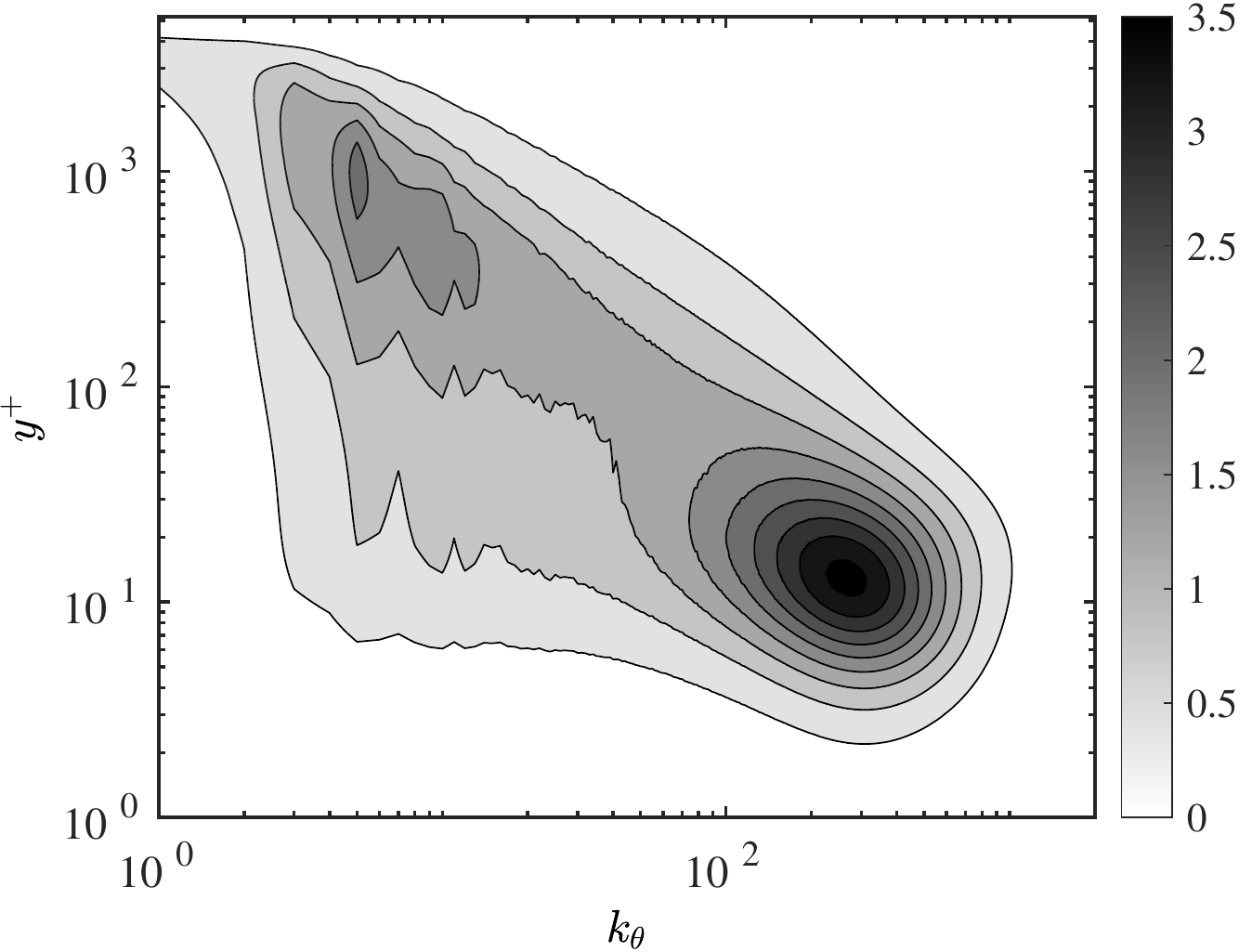}}\\
        \subfloat[]{
        \includegraphics[width=0.48\textwidth]{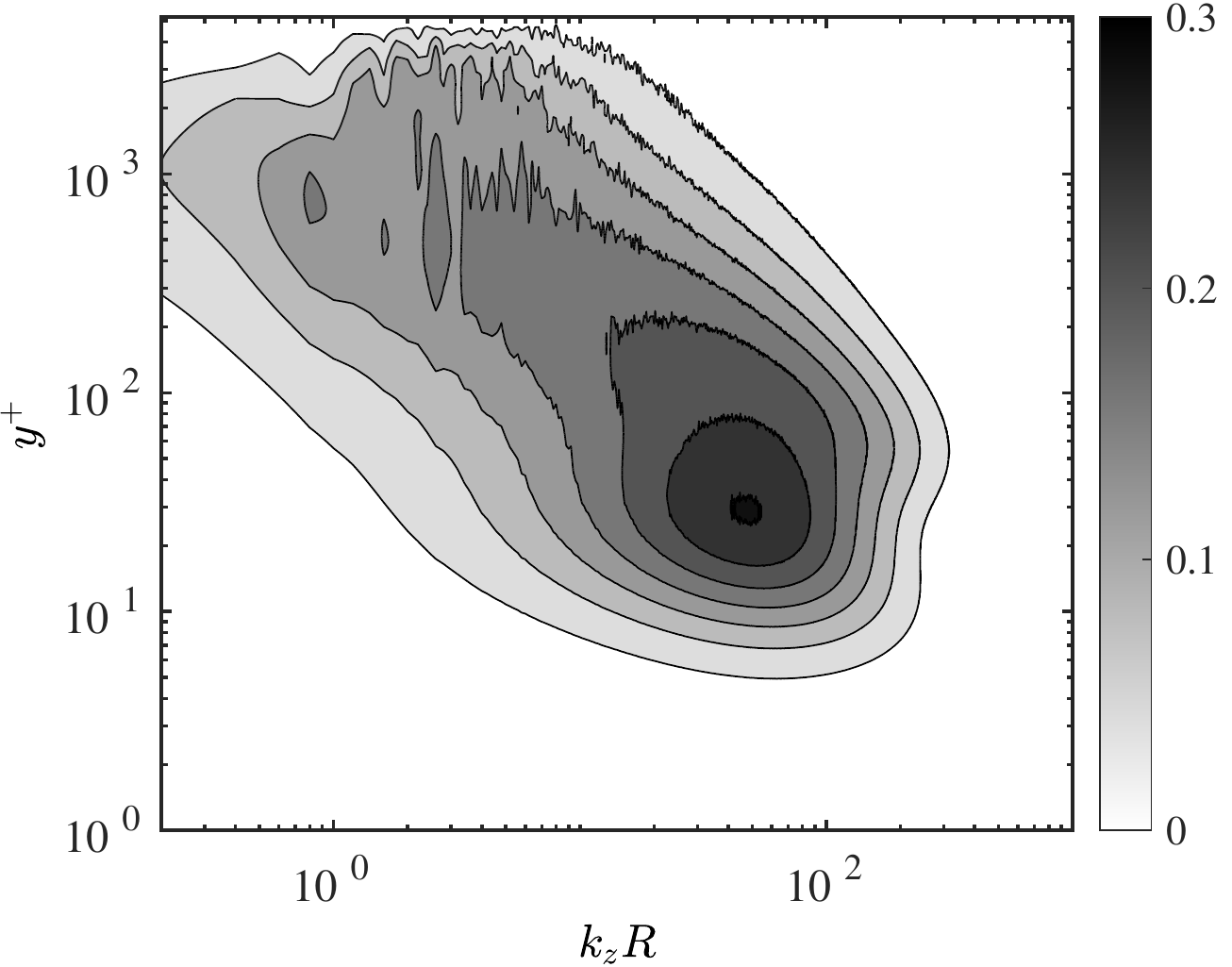}}
    \subfloat[]{
        \includegraphics[width=0.5\textwidth]{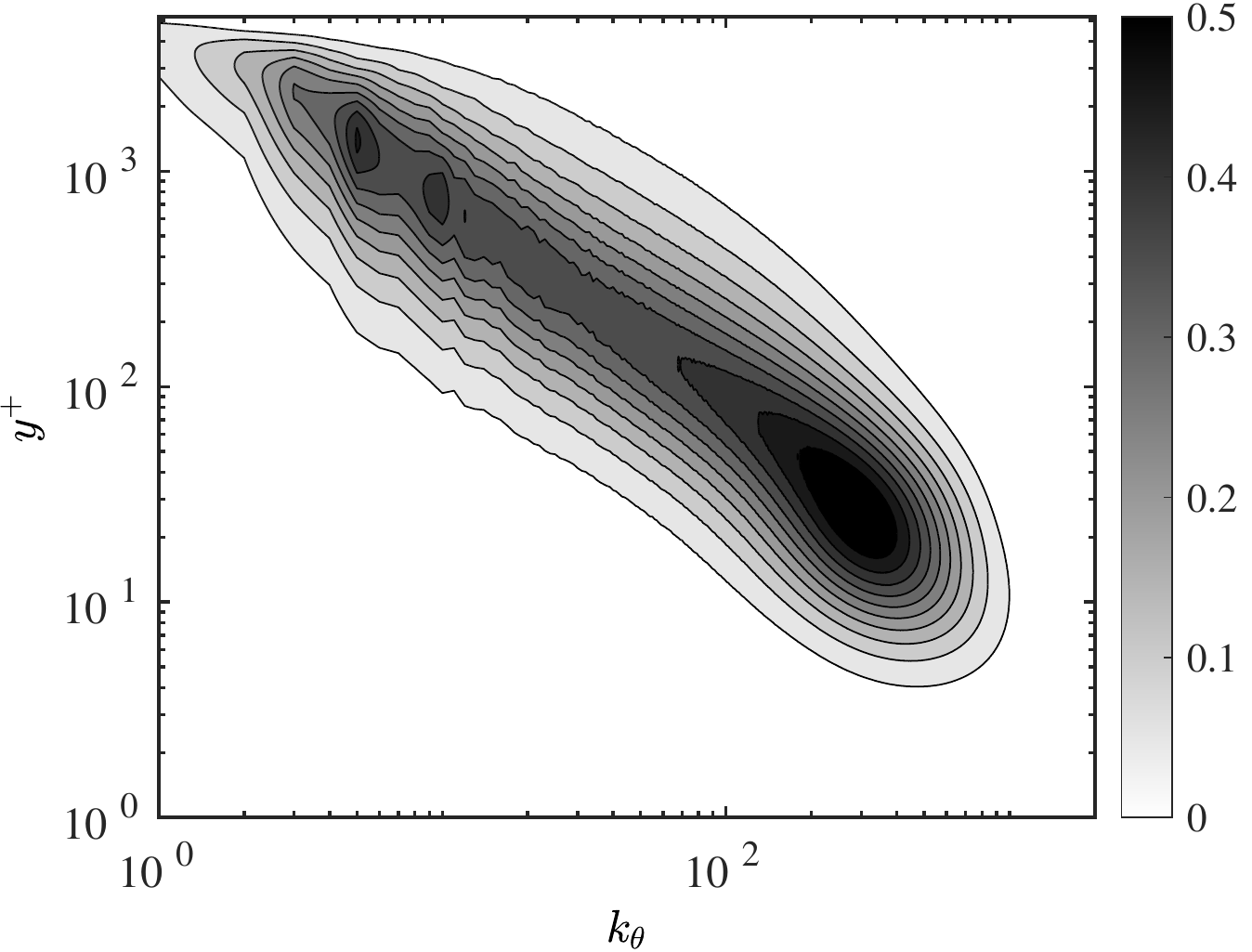}}\\
    \caption{ Wavenumber-premultiplied energy spectra for $Re_\tau=5200$: (a) $k_zE_{zz}/u^2_\tau$; (b) $k_\theta E_{zz}/u^2_\tau$; (c) $k_zE_{rz}/u^2_\tau$; and (d) $k_\theta E_{rz}/u^2_\tau$. }
    \label{fig:ek}
\end{figure}

As $Re$ increases, the separation of scales between the near-wall and outer-layer structures enlarges. In this section, the separation of scales is examined with one-dimensional velocity spectra for $Re_\tau=5200$ (figure \ref{fig:ek}).
The energy spectra of axial velocity $u_z$ in the axial direction has two distinct peaks -- the inner one located at $k_zR=40~ (\lambda^+_z=816)$ and $y^+=13$, and the outer one at $k_zR=1~ (\lambda_z=2\pi R)$, $y^+=400$.
The dual peak is more discernible in the azimuthal spectra with peaks at $k_\theta =250$ $(\lambda^+_\theta=2\pi r^+/k_\theta=131)$, $y^+=13$ and $k_\theta =6$ $(\lambda_\theta=2\pi r/k_\theta=0.846R)$, $y^+=1000$.
The $k_\theta$  of all these peaks  coincide with those found by \cite{lee2015direct} in   the channel at the same $Re_\tau$, but the physical scales are  smaller than in the channel. 
It is well known that the inner peak at $y^+=13$ is associated with the streaks that are generated through the near-wall self-sustaining cycle \citep{waleffe1997self, schoppa2002coherent}.
As  frequently seen in experiments \citep{monty2009comparison,hutchins2007evidence,rosenberg2013turbulence}, the outer peak results from  VLSMs.
The outer peak in the $\theta$-direction ($y=0.192$) was located further away than the streamwise one ($y=0.077$), which according to \cite{wu2012direct} suggests that the VLSMs in the outer region maintain their energy in $\theta$-direction  stronger than the $z$-direction . 
The pre-multiplied energy spectra of the Reynolds shear stress in  axial ($k_zE_{rz}/u^2_\tau$) and azimuthal ($k_\theta E_{rz}/u^2_\tau$) directions as a function of $y^+$  are shown in  figures \ref{fig:ek}(c) and (d), respectively. The inner peak is located at $y^+=30$ with $k_z R=49.2~ (\lambda^+_\theta=664)$ for $k_zE_{rz}/u^2_\tau$, and $k_\theta =268~ (\lambda^+_\theta=120)$ for $k_\theta E_{rz}/u^2_\tau$. 
Compared with the axial velocity spectra, although the  wavelength of the outer  peak remains identical, the magnitude is much weaker and farther away from the wall. 

\begin{figure}
    \centering

    \subfloat[]{
        \includegraphics[width=0.5\textwidth]{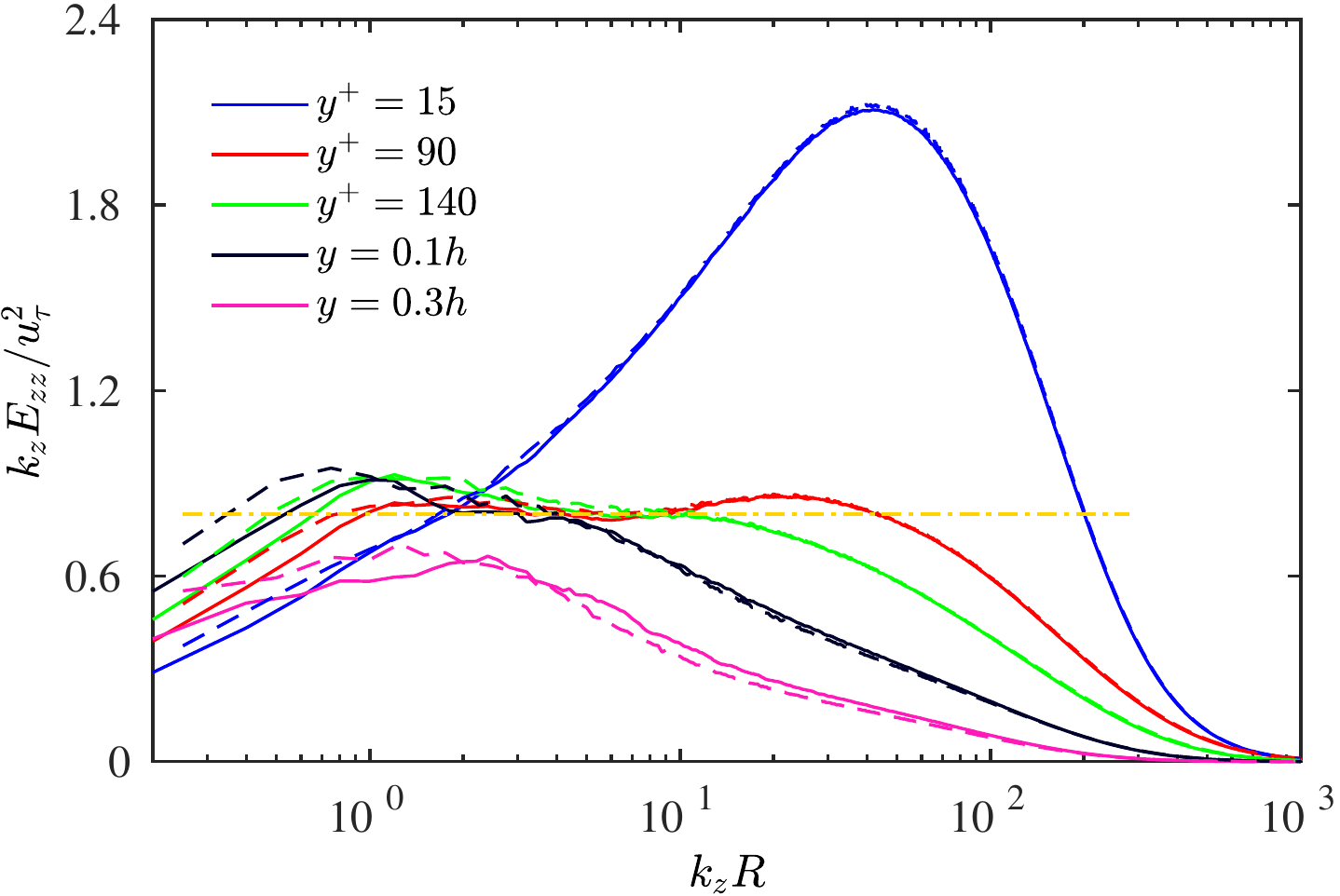}}
    \subfloat[]{
        \includegraphics[width=0.46\textwidth]{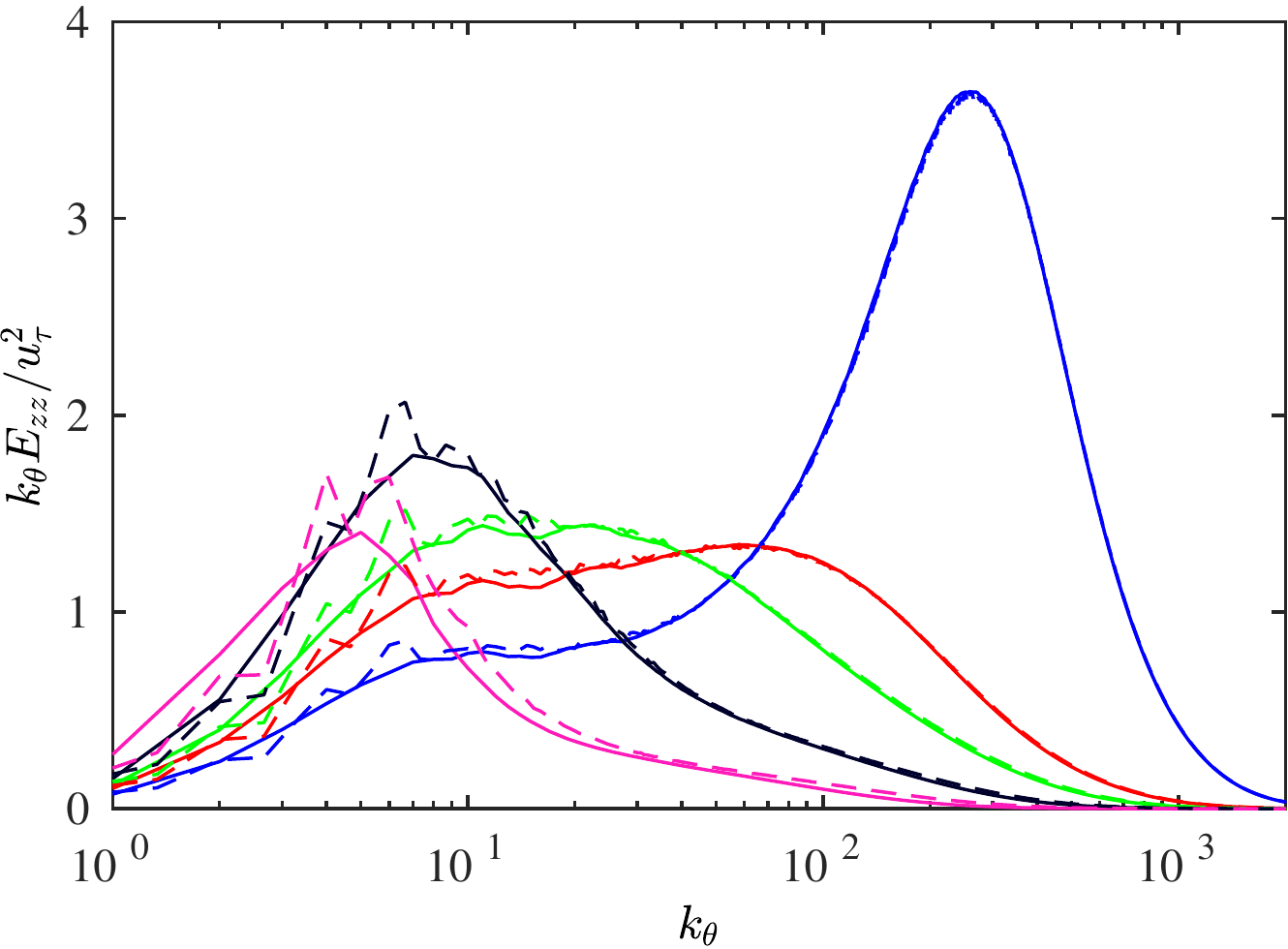}}\\

    \caption{ Comparison of the premultiplied energy spectra  between pipe (solid) and channel (dashed)  for $Re_\tau=5200$: (a) $k_zE_{zz}/u^2_\tau$ and (b) $k_\theta E_{zz}/u^2_\tau$. }
    \label{fig:ek2}
\end{figure}

Figure \ref{fig:ek2}(a) shows the one-dimensional pre-multiplied energy spectra $k_zE_{zz}/u^2_\tau$  at different $y$ locations.
For comparison, the channel data of \cite{lee2015direct} at the same $Re_\tau$ is also included.
First,  good agreement is observed  at $y^+=15$ between  channel and pipe, particularly at higher wavenumbers  -- suggesting insigniﬁcance of pipe curvature to fine-scale near-wall structures.
The scaling analysis of \cite{perry1986theoretical} suggests that  the energy spectral density of the axial velocity fluctuations $k_zE_{zz}/u^2_\tau$ should vary as $k^{-1}_z$ in the overlap region. 
The $k^{-1}_z$ region has previously been observed in the high $Re$ experiments \citep{nickels2005evidence, rosenberg2013turbulence}.
Recently,  such $k^{-1}_z$ has also been discovered in DNS of turbulent channel flow at $Re_\tau=5200$ \citep{lee2015direct}.
Similarly, a plateau  in the region  $6\le k_zR\le10$ is observed  for $90\le y^+\le170$, and the magnitude of $0.8$ agrees with experiments \citep{nickels2005evidence,rosenberg2013turbulence}.
A bimodal is observed for $y^+=90$, with the peak magnitude at low wavenumbers ($k_zR=1$) being smaller than at high wavenumber ($k_zR=30$).
Interestingly, $k_zE_{zz}/u^2_\tau$ at low wavenumbers are slightly smaller in the pipe than the channel.
Figure \ref{fig:ek2}(b) further shows the one dimensional pre-multiplied energy spectra  $k_\theta E_{zz}/u^2_\tau$  at different $y$ locations.
Again, $k_\theta E_{zz}/u^2_\tau$ agrees well between the pipe and the channel at high wavenumbers.
Consistent with those in the channel, a plateau appears for $5\le k_\theta  \le 30$  in the overlap region, with the magnitude increasing with $y^+$.
Such a plateau is present even in the viscous sublayer,  which is  the footprint of LSMs and VLSMs  near the wall \citep{mathis2009large,hwang2016inner}.  

\section{Concluding remarks}

A new direct numerical simulation providing reliable high-fidelity data  of turbulent pipe flow for $Re_\tau$ up to 5200  is presented. 
{Particular focus has been put on providing data as accurate as possible; by using a high-order numerical method, large domains and sufficient integration time with quantified uncertainty. }
The DNS is performed with a  pseudo-spectral code ``OPENPIPEFLOW'', and 
the axial extent of the domain is  $10 \pi R$ ($R$ is the pipe radius), which can be considered sufficiently long to capture all the relevant structures.
A  wealth of statistical data with uncertainty, including mean velocity, Reynolds stress and their budgets, pressure and its fluctuations, and energy spectra,  is gathered (available online at https://dataverse.tdl.org/dataverse/turbpipe).
An extensive comparison between our new pipe  data and  other simulation and experimental data is made, and   
small but still substantial and systematic differences between the various datasets   are identified.
For example, consistent lower values of the friction factor, wall shear stress fluctuations, and the inner peak of the axial velocity variance are observed for data generated using low order  methods, such as  \cite{ahn2013direct} and   \cite{pirozzoli2021one}. 
In pipe flow simulation, the only parameter apart from the $Re$ is the length of the  pipe. Once the latter is chosen large enough, all data should, in principle, be the same.
Such  discrepancies between different simulations  thus highlight the need for high-order accurate methods for this particular flow case.
{This argument is further complemented by performing an additional DNS  at $Re_\tau=2000$ with a spectral element code NEK5000, where all the statistical data generated are found to match  well the results obtained using OPENPIPEFLOW. }

Different from turbulent channel flow, the mean velocity has not yet developed a logarithmic region at $Re_\tau=5200$, yet the diagnostic function collapses well  between  our $Re_\tau=5200$ and the  channel data  of  \cite{lee2015direct} and \cite{hoyas2022wall} up to $y^+\approx250$ -- suggesting a near-wall universality of the inner scaled mean velocity.
The wall shear stress fluctuations, the inner peak of axial velocity variance, and the wall and peak of r.m.s.  pressure fluctuations continuously increase with $Re_\tau$, and  their difference between pipe and channel decreases with increasing $Re_\tau$. 
In addition, at  the $Re_\tau$ range considered,  the  $Re$ dependence  of these quantities agrees  with both the logarithmic and defect power scalings laws \citep{chen2021reynolds}.
Consistent with observations in  channel flow,  one-dimensional spectra of the axial velocity  exhibits a $k^{-1}$ dependence at  intermediate distances from the wall.

\backsection[Acknowledgements]{Computational and visualization resources provided by  Texas Tech University HPCC, TACC Lonestar,  Frontera, and Stampede2 under XSEDE  are acknowledged. Parts of the computations were enabled by resources provided by the Swedish National Infrastructure for Computing (SNIC), partially funded by the Swedish Research Council through grant agreement no. 2018-05973. This work  was partially supported by the National Science Foundation under
award number (OAC-2031650) and President's Endowed Distinguished Chair Funds. We wish to thank A. P. Willis for making the ``openpipeflow'' code open source, D. Massaro for help on the NEK5000 simulation, and  X. Wu, P. Moin, C. Chin, S. Pirozzoli,  M. Lee, L. Moser, and others for generously  sharing their  data.
}

\backsection[Declaration of interests]{The authors report no conflict of interest}

\backsection[Data availability statement]{The data that support the findings of this study are openly available  from   Texas Data Repository Dataverse at https://dataverse.tdl.org/dataverse/turbpipe.}
\appendix
\section{Estimated uncertainties for one-point statistics}\label{sec:appa} 
We briefly explain the approach employed to estimate the  uncertainties in the mean velocity and Reynolds stress components of the DNS of the pipe flow. 
During the  simulations, the time samples of the quantities contributing to statistical terms are averaged over the azimuthal ($\theta$) and axial ($z$) directions. 
To compute the central statistical moments, the temporal correlation between the spatially-averaged quantities are preserved by, for instance, writing a Reynolds stress component as $\langle u'_r u'_z\rangle = \langle \overline{u_r u_z}  -\bar{u}_r \bar{u}_z \rangle$, where overbar means averaging over~$\theta$ and~$z$.
In practice, the sample-mean estimator (SME) is used to estimate~$\langle a \rangle$ from a finite number of time-series samples~$\{a_i\}_{i=1}^n$, where $a_i=a(t_i)$ are equispaced time samples.
The SME for $\langle a \rangle$ is defined as
\begin{equation}\label{eq:sme}
    \hat{\mu}_a:=\hat{\mathbb{E}}[a]=\frac{1}{n}\sum_{i=1}^n a_i \,,
\end{equation}
where~$\hat{\mathbb{E}}[a]$ is the estimated expectation of~$a$. 
Based on the central limit theorem (CLT), for a sufficiently large number of samples, the SME converges to the true expectation via a Gaussian distribution,
\begin{equation}\label{eq:GaussMu}
    \hat{\mu}_a\sim \mathcal{N}\left(\mu_a,\sigma^2(\hat{\mu}_a) \right) \,.
\end{equation}
To estimate~$\sigma(\hat{\mu}_a)$ and hence quantify the time-averaging uncertainty in~$\hat{\mu}_a$, an analytical expression can be derived which depends on the autocorrelation of time-series~$a$ at different lags, see, e.g.~\cite{oliver:14, xavier:22} and \cite{uqRecipes}.
To avoid inaccuracy in~$\sigma(\hat{\mu}_a)$ due to the oscillations in the sample-estimated autocorrelations, especially at higher lags,  an autoregressive model is first fitted to the samples~$\{a_i\}_{i=1}^n$ which is then used to construct a smooth model for the autocorrelations. 
The details of the approach can be found in~\cite{oliver:14} and \cite{xavier:22}. 
The main hyperparameters of this UQ approach are the order of the autoregressive model and the number of training lags when modeling the autocorrelation function (ACF). 
In the present study, the optimal values of these hyperparameters are chosen based on the sampling frequency of the data at each  $Re$. 
All estimations of uncertainties have been performed using UQit~\citep{uqit}.

Following the above approach, the uncertainty in the statistical moments of any order can be accurately estimated. 
However, there are various turbulence statistics that are defined as a combination of the exponents of various moments; for instance, consider the turbulence intensity, r.m.s. fluctuations, turbulent kinetic energy, various terms in the transport equations of the Reynolds stress components, etc. 
The uncertainty in such terms can be estimated by applying the approach described in \cite{uqRecipes}. 
The main idea for estimating the uncertainty in a compound statistical term is to estimate the uncertainty in its constitutive statistical moments and also estimate the cross-covariance between the SMEs corresponding to them.  
Following this procedure, in the DNS database reported online in connection with the present study, all statistics are accompanied by an accurate estimation of the corresponding time-averaging uncertainty. 
An important aspect of this procedure is that for the statistics expressed in wall units, the uncertainty of the wall friction velocity is also taken into account. 
This means, for instance, for $\langle u'_r u'_z\rangle^+$ the uncertainty of both $\langle u'_r u'_z\rangle$ and $\langle u_\tau\rangle^2$ are considered applying a Monte Carlo-based UQ forward problem, which does not require any linearization.  

\begin{table}
    \centering
    \begin{tabular}{ccccccc}
     $Re_\tau$ & 181& 549 & 990& 2001& 5197\\
     Sampling interval ($R/U_b$) & 0.5& 0.5 &0.5&0.25 &0.25\\
     Number of samples &  6228& 1652& 1301& 1622& 683\\
    \end{tabular}
    \caption{Summary of the sampling of the flow variables used in the UQ analyses.}
    \label{tab:uqData}
\end{table}

\begin{figure}
    \centering
    \subfloat[]{
        \includegraphics[width=0.48\textwidth]{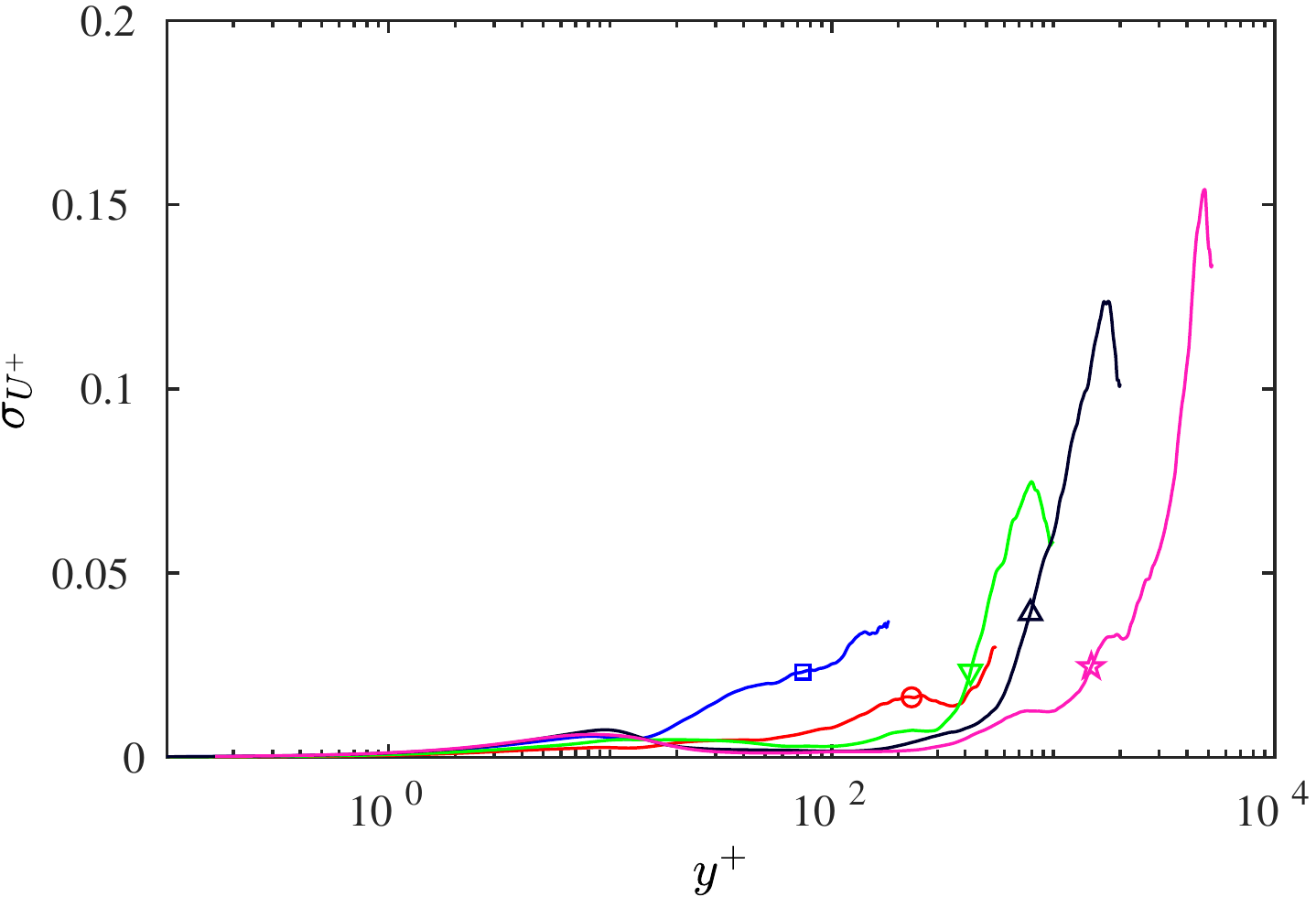}}
    \subfloat[]{
        \includegraphics[width=0.48\textwidth]{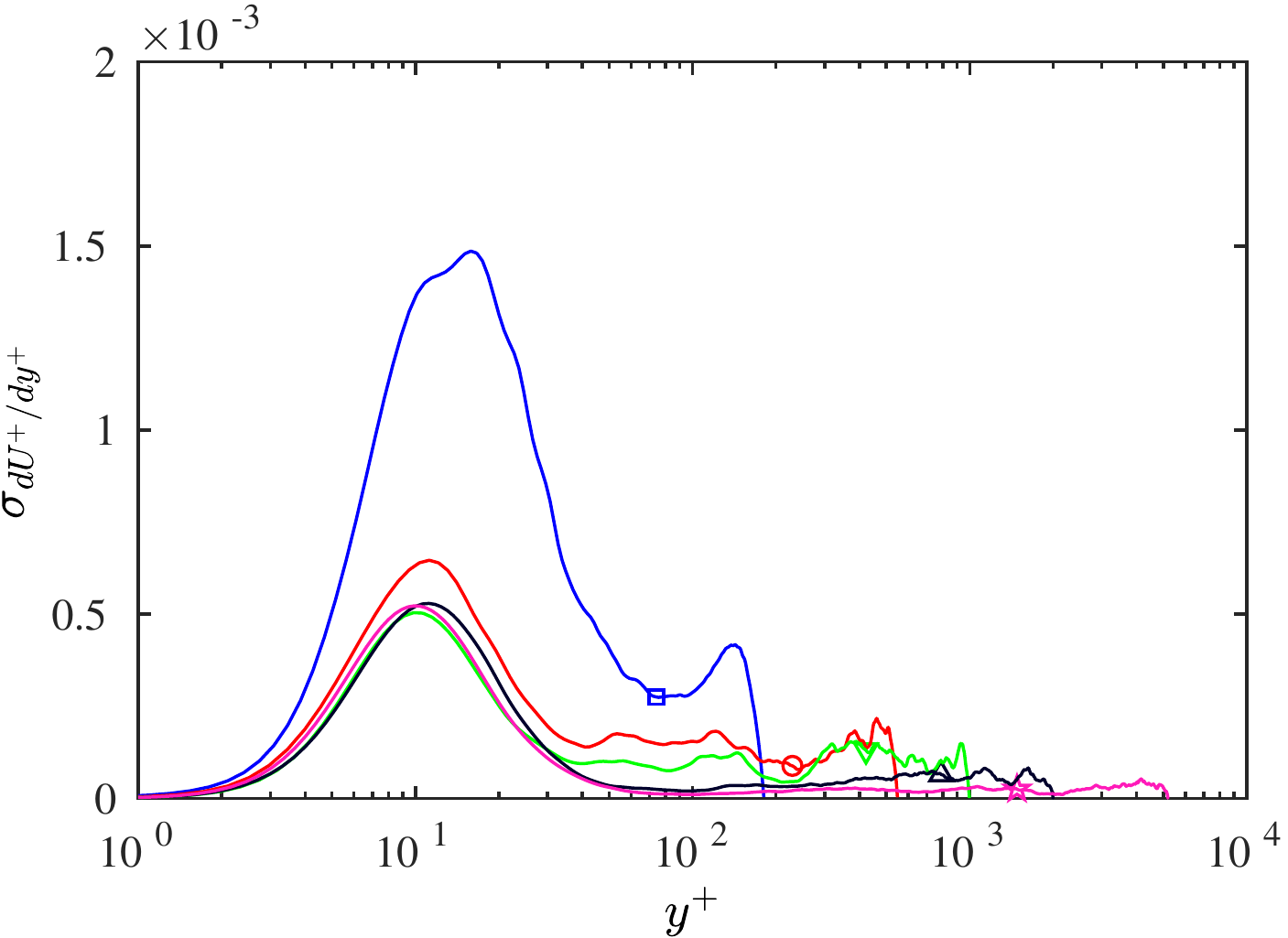}}\\
    \subfloat[]{
        \includegraphics[width=0.48\textwidth]{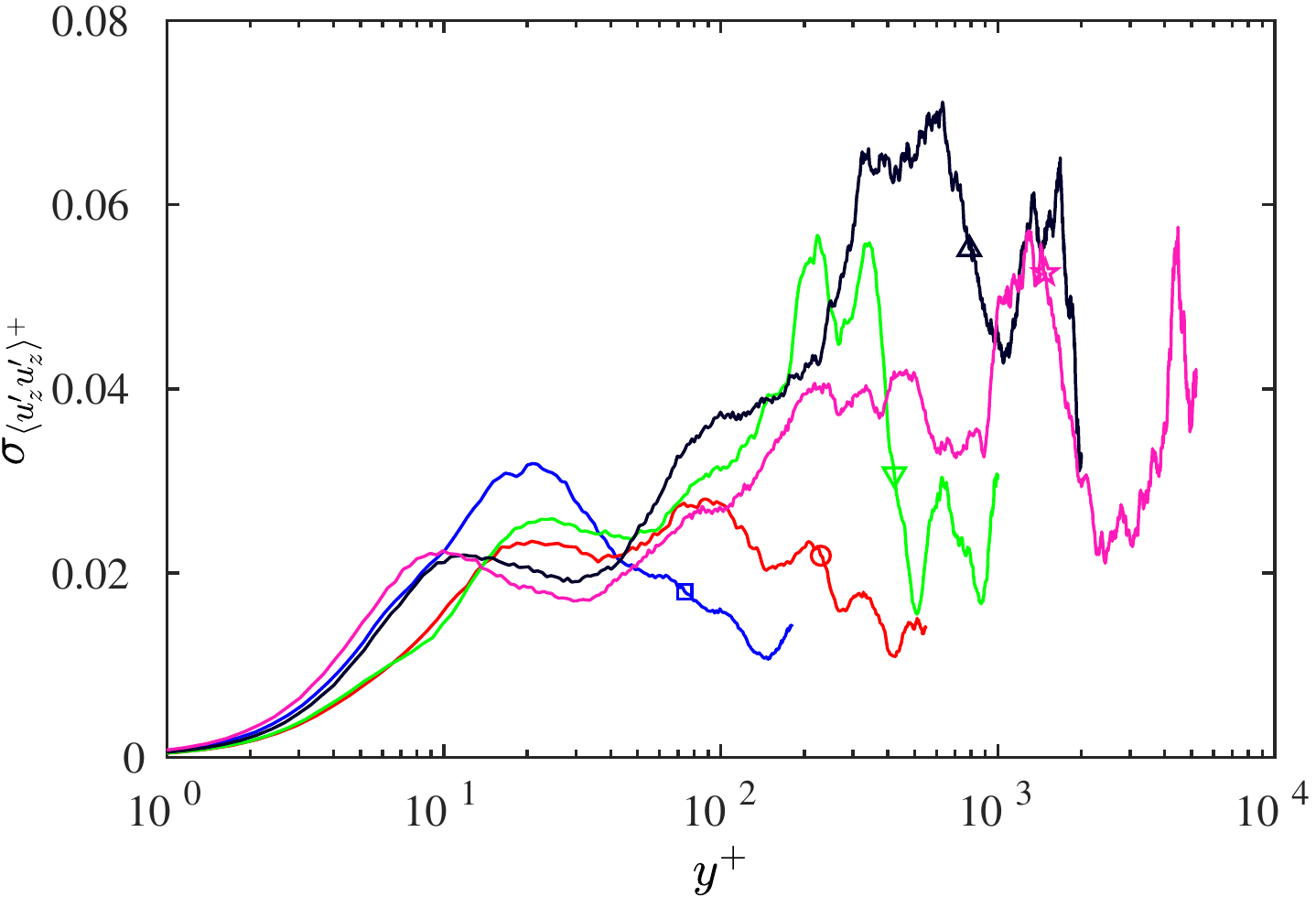}}
    \subfloat[]{
        \includegraphics[width=0.48\textwidth]{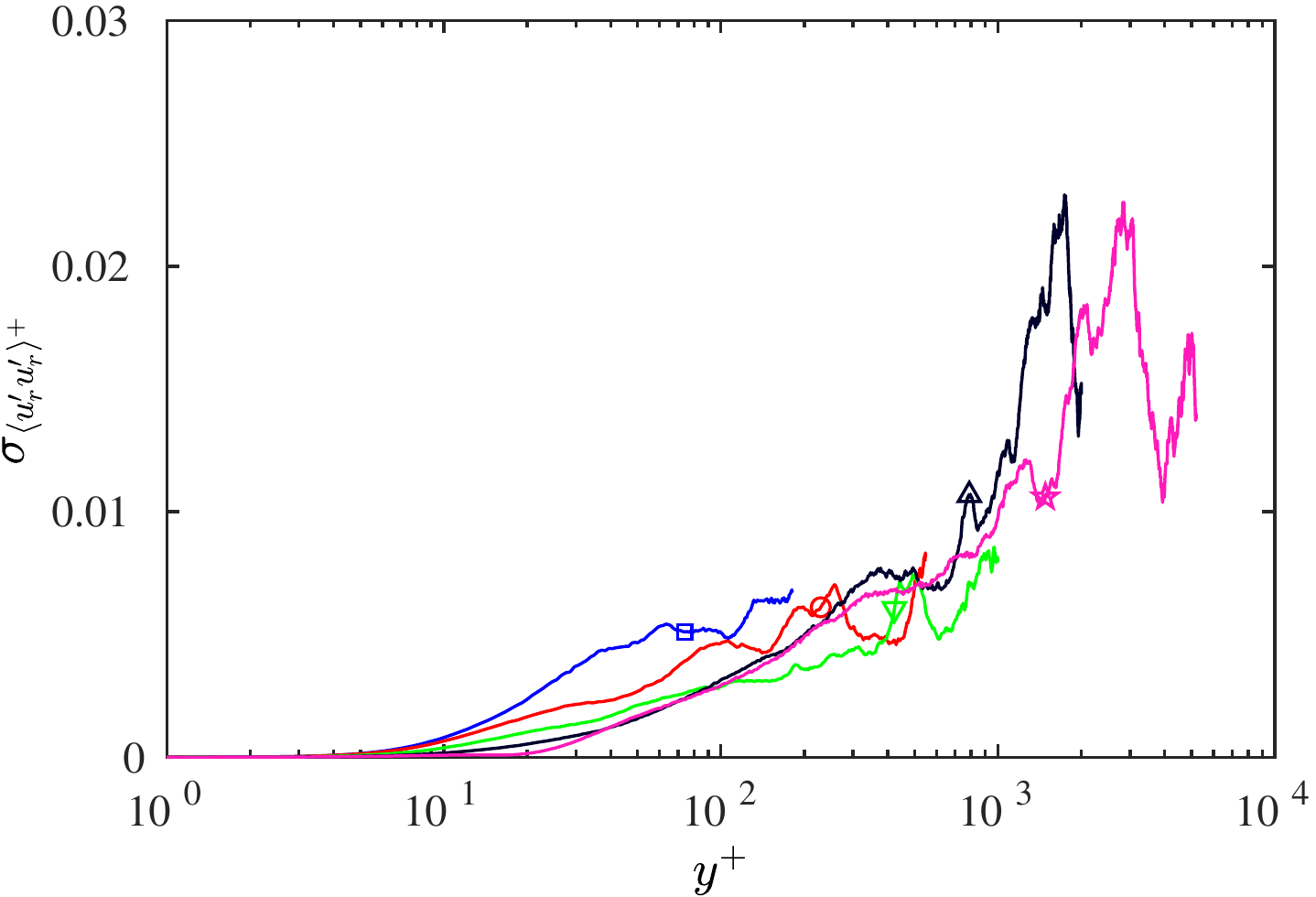}}\\
     \subfloat[]{
        \includegraphics[width=0.48\textwidth]{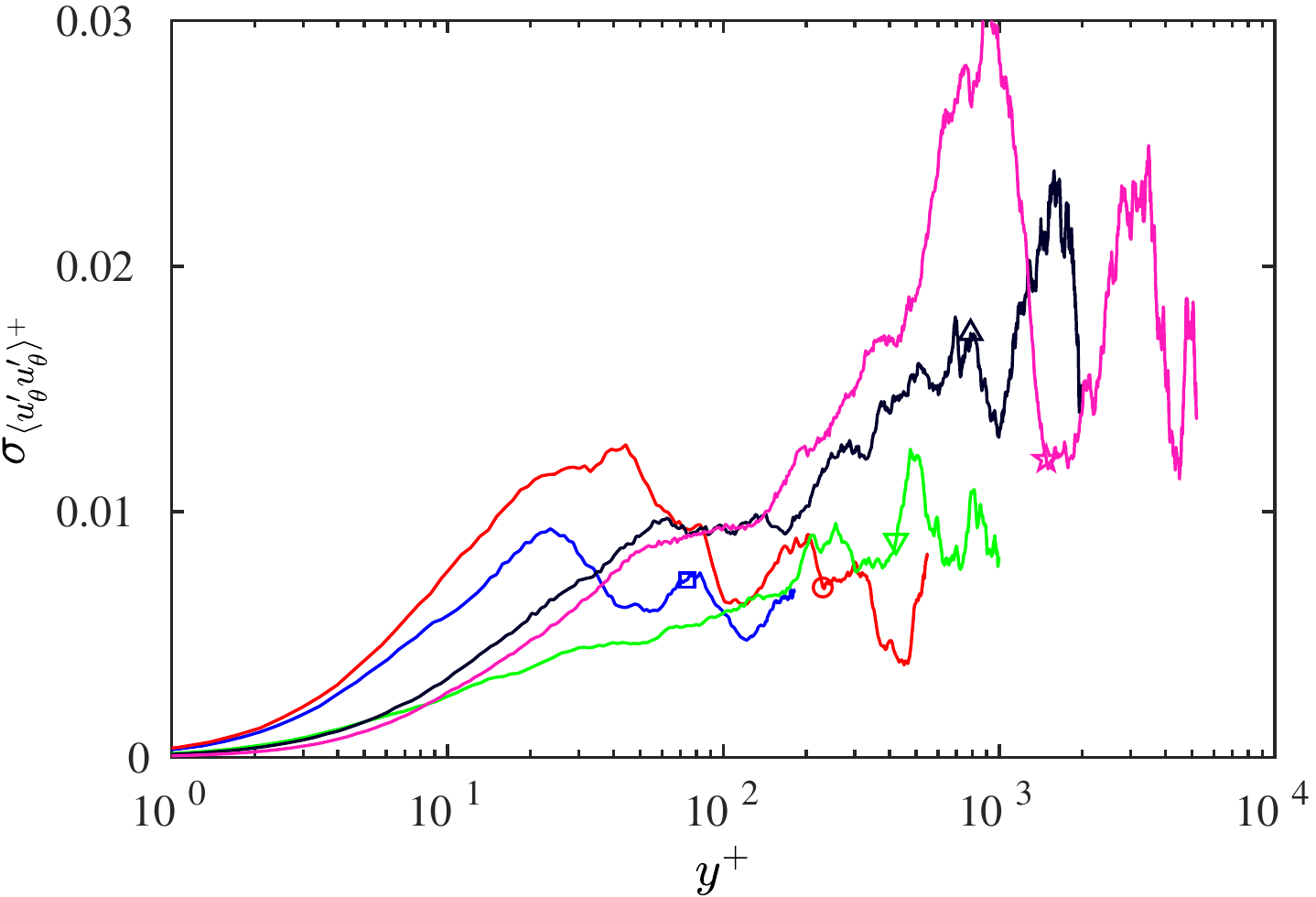}}
    \subfloat[]{
        \includegraphics[width=0.48\textwidth]{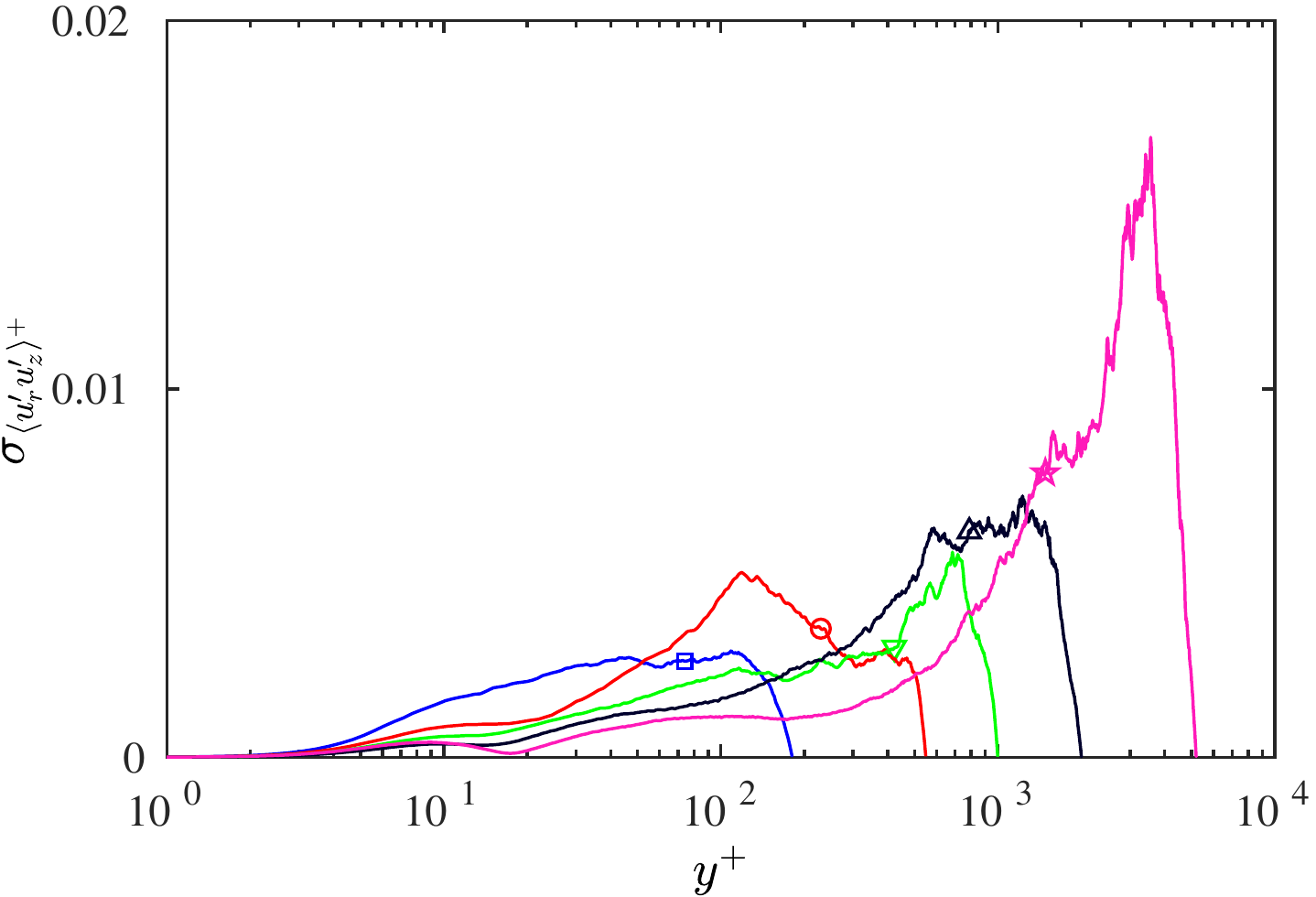}}\\
    \caption{Estimated standard deviation of different statistical quantities: (a) $U^+$; (b) $\partial U^+/\partial y^+$; (c) $\braket{u'^2_z}^+$; (d) $\braket{u'^2_r}^+$; (e) $\braket{u'^2_\theta}^+$; (f) $\braket{u'_ru'_z}^+$. }
    \label{fig:uq}
\end{figure}

Table~\ref{tab:uqData} summarizes the sampling interval and the total number of samples used for UQ for simulations at different $Re_\tau$. 
Our investigation showed that for the collected samples, an autoregressive model of order~$20$ along with the sample-estimated ACF at the first~$20$ lags, for $Re_\tau=180$, $550$, and $1000$, and $40$ lags, for $Re_\tau=2000$ and $5200$, leads to accurate models for autocorrelation of various quantities. 
For low-order moments, using sample-estimated ACF at a higher number of lags, especially  near the center of the pipe, could lead to slightly more accurate models for ACF. 
However, the difference in the resulting estimated uncertainty is below~$1\%$.

Figure~\ref{fig:uq} shows the standard deviation~$\sigma$, see (\ref{eq:GaussMu}), of the sample estimation of different inner-scaled statistical terms. 
Clearly, the estimated uncertainties vary between the moments and also in the wall-normal direction. 
However, for all quantities, the lowest uncertainty (corresponding to highest certainty) is observed near  the wall. 
Moreover, the estimated uncertainty for each quantity exhibits a similar variation in the wall-normal direction for different $Re_\tau$.

\bibliographystyle{jfm.bst}
\bibliography{fluidmechanics}

\end{document}